\documentclass[letterpaper,dvipsnames]{article}
\pdfoutput=1
\usepackage{jheppub}
\usepackage{amsmath}
\usepackage{graphicx}% Include figure files

\usepackage{array}
\usepackage{xcolor}

\usepackage{slashed}
\usepackage{afterpage}
\usepackage{appendix}
\usepackage{multirow}
\usepackage{float}
\usepackage{mathtools}
\mathtoolsset{centercolon}

\usepackage{multirow}

\newcommand{\PA}[1]{
{\color{CadetBlue}{\tt{#1}}}
}
\newcommand{\MN}[1]{
{\color{blue}\tt{Mick: #1}}
}
\newcommand{\MR}[1]{
{\color{ForestGreen}\tt{Mario: #1}}
}

\def\CP{{\sf CP}}
\def\P{{\sf P}}
\def\B{{\sf B}}
\def\L{{\sf L}}
\def\PQ{{\sf PQ}}
\def\BpL{{\sf B+L}}
\def\BmL{{\sf B-L}}
\newcommand{\TeV}{{\rm TeV}}
\newcommand{\GeV}{\ {\rm GeV}}
\newcommand{\eV}{\ {\rm eV}}

\newcommand{\ew}{{\rm ew}}
\newcommand{\EM}{{\rm em}}
\newcommand{\QCD}{{\rm QCD}}

\newcommand{\eff} {{ \rm eff}}
\newcommand{\upq}{ U(1)_{\PQ}}
\newcommand{\aGUT}{\alpha_{\rm GUT}}
\newcommand{\GGUT}{\mathcal{G}}
\newcommand{\GUT}{{\mathbb{G}}}
\newcommand{\MGUT}{M_{\rm GUT}}

\title{Axion Couplings in Grand Unified Theories}

\author{Prateek Agrawal,}
\author{Michael Nee,}
\author{and Mario Reig}

\affiliation{Rudolf Peierls Centre for Theoretical Physics, 
University of Oxford, Parks Road, Oxford OX1 3PU, United Kingdom}
\emailAdd{prateek.agrawal@physics.ox.ac.uk}
\emailAdd{michael.nee@physics.ox.ac.uk}
\emailAdd{mario.reiglopez@physics.ox.ac.uk}

\abstract{
We show that the couplings of axions to gauge bosons are highly restricted in Grand Unified Theories where the standard model is embedded in a simple 4D gauge group. The topological nature of these couplings allows them to be matched from the UV to the IR, and the ratio of the anomaly with photons and gluons for any axion is fixed by unification. This implies that there is a single axion, the QCD axion, with an anomalous coupling to photons. Other light axion-like particles can couple to photons by mixing through the QCD axion portal and lie to the right of the QCD line in the mass-coupling plane. Axions which break the unification relation between gluon and photon couplings are necessarily charged under the GUT gauge group and become heavy from perturbative mass contributions. A discovery of an axion to the left of the QCD line can rule out simple Grand Unified models. Axion searches are therefore tabletop and astrophysical probes of Grand Unification.
}

\begin{document}

\maketitle

\section{Introduction}

\label{sec:introduction}

Topological couplings provide a unique opportunity to connect the far UV and IR dynamics of theories. In general, even if the UV theory is simple, the IR dynamics can be very complicated, as exemplified by the rich structure of nuclear physics that arises from the relatively simple QCD Lagrangian. Topological couplings, however, are unaffected by such dynamics and are invariant under renormalisation group flow. If the UV theory is simple then this provides restrictive constraints on the topological interactions in the IR with direct experimental consequences.
The axion-photon coupling is extremely important for experimental searches (see~\cite{Arias:2012az,Graham:2015ouw,Bauer:2017ris,Irastorza:2018dyq} for a review) and gets a UV contribution from a quantised anomaly coefficient which is a topological quantity~\cite{'tHooft:1979bh}. 
There are additional mixing contributions to the coupling that come from working in a canonical basis of mass eigenstates, but this mixing is calculable in the IR given a mass generation mechanism for the axion~\cite{Weinberg:1977ma,Wilczek:1977pj}. In this paper we study the restrictions on axion-photon couplings in Grand Unified Theories (GUTs). The axion-photon coupling is particularly relevant as the majority of experimental searches for axions aim to detect axions through this coupling~\cite{Wilczek:1987mv,Sikivie:1983ip,Sikivie:1985yu,Dreyling-Eschweiler:2014mxa,TheMADMAXWorkingGroup:2016hpc,Andriamonje:2007ew,Anastassopoulos:2017ftl,Asztalos:2009yp,Armengaud:2014gea,Majorovits:2016yvk,Kahn:2016aff,Arvanitaki:2017nhi,Baryakhtar:2018doz,Chaudhuri:2018rqn}.

Axions are compact bosons with a discrete gauged shift symmetry. In this context it is very useful to interpret axions as ``0-form'' gauge fields, and the shift symmetry as a large gauge transformation. This language highlights the topological nature of axions. 
The axion couplings to gauge fields are Chern-Simons like couplings which are topological quantities, and we show in section~\ref{sec:quantisation} that these couplings are quantised.   Axions are compelling new physics candidates, motivated from the bottom-up perspective by their role in solving the strong $\CP$  problem~\cite{Peccei:1977hh,Weinberg:1977ma,Wilczek:1977pj,Kim:1979if,Shifman:1979if, Dine:1981rt,Zhitnitsky:1980tq} and dark matter~\cite{Preskill:1982cy,Abbott:1982af,Dine:1982ah}, as well as from top-down constructions in string theory where they are ubiquitously present~\cite{Svrcek:2006yi,Arvanitaki:2009fg,Cicoli:2012sz}.

The QCD axion~\cite{Weinberg:1977ma,Wilczek:1977pj} is a particularly well-motivated axion, which couples anomalously to QCD. If QCD instantons dominate over the other contributions to the axion potential, then by the Vafa-Witten theorem~\cite{Vafa:1984xg} the axion potential is minimised when the $\CP$ violating phase $\bar{\theta}$ is set to zero, dynamically solving the strong $\CP$ problem. This also leads to a relationship between the mass of the QCD axion and its coupling to gauge bosons. In particular the coupling of the QCD axion to photons is~\cite{diCortona:2015ldu} 
\begin{align}
    &\frac{g_{a \gamma \gamma}}{m_a} =
    \frac{\alpha_{\EM}}{2\pi}
    \frac{\sqrt{z}+\frac{1}{\sqrt{z}}}{ m_\pi f_\pi}
    \left(\frac{E}{N}-1.92\right),
    \qquad
    z = \left(m_u/m_d\right).
    \label{eq:QCDline1}
\end{align}
The rational number $E/N$ which represents the ratio of the anomaly coefficients to photons and gluons. This line in $g_{a \gamma \gamma}$--$ m_a$ space is known as the QCD line. For the purposes of experiments which search for the axions via the anomalous photon coupling, the QCD axion is effectively a one parameter model.

In addition to the QCD axion there could be other (ultra)light axions that can couple to photons, referred to as axion-like particles (ALPs). ALPs generally do not have low-energy instanton contributions as in the case of QCD axion, and the dominant contribution to their mass arises from the breaking of their continuous shift symmetry by UV effects \cite{Hui:2016ltb}. Such effects are thought to be unavoidable in theories of quantum gravity~\cite{Banks:2010zn}, but they can be exponentially small~\cite{Kallosh:1995hi}. %Mehta:2020kwu,Mehta:2021pwf Therefore, their mass $m_a$ and coupling to photons $g_{a\gamma\gamma}$ are treated as independent. 
ALPs may therefore have much smaller masses than the QCD axion and have a coupling to photons which is independent of their mass.
%This has motivated a broad search for ALPs which couple to photons off the QCD line that do not couple to gluons~\cite{Agrawal:2021dbo}. 
This has motivated a broad search for ALPs with photon coupling and mass off the QCD line~\cite{Agrawal:2021dbo}, often in the low-mass region of parameter space. %In the context of GUTs, we show that the coupling of these extra axions to photons is generated by mechanisms that are correlated with the mass for the axion, highlighting specific targets in the $m_a$--$g_{a\gamma\gamma}$ space.
We show that if the Standard Model (SM) is unified in the UV, however, the ALPs coupling to photons is generated by mechanisms that are correlated with the mass for the axion and points to a preferred region of ALP parameter space where $g_{a \gamma \gamma} /m_a$ is smaller than the QCD axion expectation.

%Grand Unification remains one of the most compelling UV completions that explains some of the features of the SM. In this work we consider theories where the SM gauge fields are embedded in a 4D gauge theory with a simple group at some high energy scale~\cite{Georgi:1974sy,Fritzsch:1974nn}.
Grand Unification remains one of the most compelling UV completions that explains some of the features of the SM by embedding the SM gauge fields in a simple group at some high energy scale~\cite{Georgi:1974sy,Fritzsch:1974nn}. %The simple gauge group is responsible for the most compelling predictions of grand unification -- the explanation of SM quantum numbers and the relative size of the gauge couplings, and as we show in this paper, axion phenomenology. 
This framework provides an explanation for the SM quantum numbers and the relative size of the gauge couplings in the IR. As we will show in this paper, it also has strong implications for axion phenomenology. The underlying reason for this is that the couplings of axions to gauge bosons are determined by anomaly coefficients, which are necessarily quantised and dictated by gauge quantum numbers in the UV. 
%Therefore, it is a natural expectation that GUTs give strong predictions for axion phenomenology. 

The requirement that all of the SM gauge groups unify in the UV requires that the axion must couple anomalously to \emph{all} gauge bosons or none at all. %Unification of the SM gauge groups in the UV requires that the anomaly coefficients which determine the axion coupling to photons and to gluons are therefore related -- in particular it is not possible for an ALP to couple to photons without coupling to gluons.
In the limit where mass mixing between axions can be ignored there is a single axion -- the QCD axion -- which couples to photons as well as gluons, and all other ALPs are fully decoupled (barring derivative couplings to fermions, as we discuss below). The QCD axion anomalously coupled to the GUT gauge group follows equation~\eqref{eq:QCDline1} with the ratio $E/N$ fixed by the structure of the GUT gauge group\footnote{The minimal GUT prediction $E/N=8/3$ was noted long ago \cite{Srednicki:1985xd}. The anomaly ratio can take more general values in GUT theories where the SM is non-trivially embedded. See Appendix \ref{sec:embedding}.}. We discuss the GUT prediction for the phenomenology of the QCD axion in section~\ref{sec:GUTQCDaxion}.

%In GUTs with a simple gauge group axions which do not have an anomaly with QCD will also not couple anomalously to photons.
Any axion with mass and coupling not on the QCD line we refer to as an ALP. Once effects of mass mixing are considered an ALP can inherit a coupling to the photon through the `QCD axion portal'. In this case, in the absence of tuning, a light ALP $b$ will have couplings to gauge bosons which are suppressed by $m_b^2 / m_{\QCD}^2$. Thus we see that through a completely different mechanism, an ALP in these theories still has a correlation between its mass and photon coupling. If an ALP does not couple through mixing with the QCD axion, it can pick up a coupling to the photon if it is itself charged under the GUT gauge group (but neutral under the surviving SM group), as is the case in composite axion models~\cite{Kim:1984pt,Choi:1985cb,Kaplan:1985dv,Randall:1992ut}. In this case it will pick up a perturbative mass from GUT interactions, analogous to the electromagnetic contribution to the charged pion masses. %In section~\ref{sec:GUTALP} we study the effects of mass mixing and kinetic mixing of axions, axions charged under the GUT gauge group and the implications of additional dark photons mixing with the SM photon. 
In section~\ref{sec:GUTALP} we determine the photon coupling of ALPs generated through each of these two effects as well as the implications of additional dark photons mixing with the SM photon. We show that the coupling of ALPs to photons is necessarily weaker than that of the QCD axion for the same mass, in the parameter space to the right of the QCD line.

ALP-fermion couplings are not quantised and are therefore not suppressed in GUTs, implying that experiments that search for ALPs through their fermion couplings~\cite{Budker:2013hfa,Arvanitaki:2014dfa} may be promising methods to search for ALPs in the context of GUTs, which are discussed in section~\ref{sec:ALP_pheno}. Assuming flavour-conserving couplings, the ALP coupling to electrons in astrophysical systems turns out to give the most stringent bounds on the ALP parameter space. Flavour-violating couplings, if present, can also place strong constraints on the ALP decay constant in laboratory experiments~\cite{Wilczek:1982rv, MartinCamalich:2020dfe, Bauer:2021mvw}. We show that, remarkably, while axion-mediated forces are unsuppressed for the QCD axion, for light ALPs the monopole-dipole or the monopole-monopole interaction is strongly suppressed. The experimentally interesting monopole-dipole forces can distinguish between the QCD axion and ALPs in GUTs.

The main result our work highlights is that the phenomenology of the QCD axion and additional ALPs is strongly constrained in GUT theories. Axions therefore give us a low-energy handle to probe grand unification in table-top experiments~\cite{Arvanitaki:2014dfa,Raffelt:2012sp,Okawa:2021fto} and in the sky~\cite{Arvanitaki:2009fg, Agrawal:2019lkr}. The traditional experimental strategy to look for GUTs is to search for the decay of protons into mesons and anti-leptons. The current limit on the life-time of the proton: $\tau > 2.4\times 10^{34}$ yr~\cite{Super-Kamiokande:2020wjk} constrains the GUT scale to be $M_{\rm GUT} \gtrsim 2\times 10^{16}$~GeV. The bound on the life-time is expected to be improved by a factor of around 10 at Hyper-K \cite{Abe:2011ts}, which results in only a factor of $\sim 2$ in the reach for $M_{\rm GUT}$. Our results imply that a discovery of a light ALP with $g_{a \gamma \gamma}/m_a$ larger than the QCD prediction can potentially determine if the SM gauge groups are unified in the UV. ALP searches can then give us a novel experimental handle on GUTs.

In the literature the term GUT applies more widely to theories which only partially unify the SM gauge groups. We discuss these theories in section~\ref{sec:motorbikes} and find that the possibility of finding a light ALP in these theories is correlated with sacrificing some of the predictions of GUTs -- namely, charge quantisation, the prediction of the weak mixing angle and non-existence of exotic fractionally-charged states -- although this is somewhat model dependent. %An important class of grand unified theories in string theory where the unification and compactification have interesting interplay do not fall in the category of theories considered here. These theories, known as orbifold GUTs~\cite{Kawamura:1999nj,Hebecker:2001wq,Hall:2001pg,Altarelli:2001qj}, and related string theoretical scenarios will be studied in more detail in upcoming work~\cite{second_paper}.
Another important class of related theories are GUTs in higher dimensions, where unification and compactification of extra dimensions have interesting interplay. These theories, known as orbifold GUTs~\cite{Kawamura:1999nj,Hebecker:2001wq,Hall:2001pg,Altarelli:2001qj,Hall:2001tn,Hebecker:2001jb} and related string theoretic scenarios will be studied in more detail in upcoming work.

It is worth comparing the results in this work with previous literature on axions and GUTs, which has focused on the dynamical aspects of unification such as $\beta$ functions, matter representations, the GUT scale and axion mass. In~\cite{Giudice:2012zp}, the emphasis was on possible representations of the SM that can lead to perturbative unification. In the models of~\cite{Georgi:1981pu,Nilles:1981py,Wise:1981ry,DiLuzio:2018gqe,Ernst:2018bib,FileviezPerez:2019fku}, the GUT scale was tied to the axion decay constant\footnote{See also \cite{Davidson:1983fy,Davidson:1983fe,Chen:2021haa} for connections of PQ symmetry to GUTs in a different context.}. Our work focuses on the topological aspect of axion couplings, and is directly tied to charge quantisation in GUTs. Thus it is complementary to the dynamical details of the theory such as the scale of $\PQ$ breaking or patterns of GUT symmetry breaking. %In particular, our results are independent of matter representations or the scale of $\PQ$ breaking relative to the GUT scale.

\section{Quantisation of axion couplings}

\label{sec:quantisation}

The central idea underlying our analysis is the fact that axion couplings to gauge bosons are quantised, and can be anomaly matched between the UV and the IR~\cite{Agrawal:2017cmd,Fraser:2019ojt}. In this section we review the arguments that show that the couplings of an axion to a gauge boson are quantised\footnote{In~\cite{Sokolov:2022fvs} the question of quantisation of axion couplings with photons is revisited, however the coupling to non-Abelian gauge bosons is still assumed to be quantised. We will not have anything to add to the discussion in this paper.}. The effects of possible kinetic and mass mixing will be studied in subsequent sections and form the bulk of our analysis which shows that the coupling and mass for ALPs are correlated in unified theories.

The axion is defined as a compact scalar $a$, or equivalently a pseudoscalar field with a discrete gauge symmetry $a \simeq a + 2\pi F_a$. In many examples, the compactness of the axion follows from its role as the Goldstone boson when a compact global Peccei-Quinn symmetry $\upq$ is spontaneously broken. In other cases it arises from dimensional reduction of a gauge theory with a compact gauge group. It is useful to think of the discrete shift of the axion as a large gauge transformation for a 0-form gauge field. The quantisation of axion couplings then follows similar logic to the quantisation of $U(1)$ gauge charges.

Anticipating our discussion of GUTs, we start with an example of a single axion coupled to a gauge field $\GGUT$ associated with a simple group $\GUT$, and review the 
quantisation of axion couplings to gauge groups.
The axion-photon coupling is intimately related to the Wess-Zumino-Witten term in the chiral Lagrangian, or the Chern-Simons couplings of gauge bosons in odd-dimensions. The Lagrangian describing an axion anomalously coupled to the gauge group $\GUT$ is
\begin{align}\label{eq:axion_Lag}
    \mathcal{L}
    &=
    -\frac{1}{4} \GGUT^a_{\mu\nu} \GGUT^{\mu \nu, a}
    +\frac12 \partial_\mu a \partial^\mu a
    +\mathcal{A}  \frac{a}{F_a} \frac{\aGUT}{8\pi}{\GGUT}^a_{\mu\nu}
    \widetilde{\GGUT}^{\mu\nu,a}
\end{align}
where $\widetilde{\GGUT}^{\mu\nu} = \frac12 \epsilon^{\mu\nu\alpha\beta} \GGUT_{\alpha\beta}$.
The parameters $\alpha$ and $F_a$ are not topological and are sensitive to geometry and dynamical effects such as RG running. The anomaly coefficient, $\mathcal{A}$, however, is quantised and by the usual arguments of anomaly matching transmits information from the UV to the IR unpolluted by intervening physics\footnote{Equation~\eqref{eq:axion_Lag} is defined in a basis where the determinant of the fermion mass matrix is real. It can be defined in more physical terms using the axion to photon decay amplitude, see discussion in~\cite{Baryakhtar:2013wy}.}. In this section we review one general argument that shows this quantisation of the axion-photon coupling.

A concise way to see the quantisation of the axion photon coupling is by analogy with Chern-Simons theory in odd dimensions. If we put the 4D theory in Euclidean space on $S_4$ by adding a point at infinity to $\mathbb{R}_4$, on a constant axion background the action is proportional to the winding number of the gauge field configuration
\begin{align}
    \int_{S_4}
    \frac{\aGUT}{8\pi}{\GGUT}^a_{\mu\nu}
    \widetilde{\GGUT}^{\mu\nu,a}
    = n
    \in \mathbb{Z}
    .
\end{align}
The hallmark of a Chern-Simons action is that it is not gauge-invariant (in this case under the discrete gauge symmetry of the axion, $a \to a + 2\pi F_a$), 
\begin{align}
    I[a+2\pi F_a]
    &=
    I[a]+
    2\pi n \mathcal{A}
    \,.
\end{align}
However, it is gauge invariant modulo $2\pi$, and hence can be used as a quantum action, $\exp(i I)$ in the path integral for $\mathcal{A} \in \mathbb{Z}$. %A violation of the axion discrete shift symmetry indicates that the field is non-compact, 

The discussion above can easily be generalised to include multiple massless axions. If there is a set of global $U(1)$ symmetries anomalous\footnote{Technically these global symmetries are broken since the associated current has an ABJ anomaly with a dynamical gauge field. We will continue to use the term anomalous for such symmetries, since we are interested in matching the anomalies in weakly coupled regimes for the SM gauge fields. Similarly, breaking of the global $U(1)$ symmetries by quantum gravitational effects will be negligible for the cases of interest.} under $\GUT$ that are realized non-linearly by axions, the $U(1)$ symmetries can always be redefined so that only one of them is anomalous under $\GUT$. We will call this symmetry $\upq$. This sort of redefinition is very similar to the case of the baryon and lepton numbers in the SM. The $U(1)_\B$ and $U(1)_\L$ symmetries are each anomalous with respect to $SU(2)_{\ew}$, but can be redefined into $U(1)_{\BmL}$ and $U(1)_{\BpL}$ symmetries, of which only $U(1)_{\BpL}$ is anomalous. In a similar way, we can choose a linear combination of the $U(1)$ symmetries such that only $\upq$ will be anomalous under $\GUT$. When $\upq$ is spontaneously broken the corresponding axion will saturate the anomaly and other axions will remain decoupled from gauge fields. This is consistent with our discussion above of quantisation of couplings -- only one axion $a$ couples to $\GGUT$ with a non-zero integer $\mathcal{A} \in \mathbb{Z}$, and all other axions $b_i$ couple with the integer $\mathcal{A}_{b_i}=0$.

This is a rather strong statement, so it is worth emphasizing the assumptions under which this is true.
\begin{itemize}
    \item We have not included the effects of axion mass terms or kinetic mixing, which can generate mixing between axions. When these effects are included an ALP $b$ with vanishing anomaly coefficient may couple to $\GGUT \tilde \GGUT$ through mixing with the state $a$. %We will see that we can get definite predictions for such ALPs which couple through the QCD axion portal.
    % We have assumed that apart from the mixed anomaly with $\mathbb{G}$, all the $U(1)$ symmetries are exact. In general, all axions are expected to pick up a mass from various instanton effects in quantum gravity. For there to be a solution to the strong $\CP$ problem, there needs to be a set of axions for which the QCD instantons vastly dominate the potential. In this case there is one QCD axion, $a$ and the other axions $b_i$ only couple to the SM through small mixing effects.
    \item The axion is assumed to be neutral under the full GUT gauge group $\GUT$. If an axion is charged under $\GUT$ there may be multiple axions coupled to photons. The GUT charge however implies that the axion picks up a perturbative mass from GUT dynamics. 
    \item In our analysis in this paper, we assume a simple unified gauge group $\GUT$. We study the implications of relaxing this assumption in section~\ref{sec:motorbikes} within 4D field theories. The interplay of axions and grand unification in higher dimensional GUTs will be studied in a follow-up paper. %~\cite{second_paper}.
%    \item \PA{What happens if the SM gauge group itself is emergent?}
\end{itemize}
In each of these cases, we find that the ALP mass and coupling to photons are correlated, and in specific cases lead to concrete targets in the $m_a$--$g_{a\gamma\gamma}$ plane. In particular, we find that
\begin{align}
   \frac{g_{b \gamma \gamma}}{m_b} &\gg  \frac{1}{m_\pi f_\pi} \,   \label{eq:lightALP}
\end{align}
is not possible in 4D GUT theories unless there is very finely tuned cancellation in contributions to the axion mass.

\section{GUT predictions for the QCD axion}
\label{sec:GUTQCDaxion}

In this section we review the predictions for the couplings of the axion $a$ anomalously coupled to a simple GUT group $\GUT$ in the UV which contains the SM. This corresponds to the well-known case of the QCD axion which solves the strong $\CP$ problem. 
After spontaneous symmetry breaking (SSB) of $\GUT$ and $SU(2)_{\ew}$ the axion will have couplings to gauge bosons given by: 
\begin{align}
     \mathcal{A}
     \frac{\aGUT}{8\pi}\frac{a}{F_a}\GGUT_{\mu\nu}\widetilde{\GGUT}^{\mu\nu}
    \rightarrow \frac{a}{F_a}\left[ 
    \frac{\alpha_{\EM}}{8\pi}E\,F_{\EM,\mu\nu}\widetilde{F}_{\EM}^{\mu\nu}
    + \frac{\alpha_{s}}{8\pi}N\,G^a_{\QCD, \mu\nu}\widetilde{G}^{a,\mu\nu}_{\QCD}\right] \, .
    %+ c_{h}G^\prime\tilde{G}^\prime+...\right] \, .
    \label{eq:aGGcouplings}
\end{align}
The anomaly coefficients $E$ and $N$ are rational numbers that set the axion coupling to photons and gluons respectively and are fixed by the embedding of the SM into $\GUT$. In the low-energy theory the Lagrangian is parameterised as
\begin{align}
    \mathcal{L}
    &=
    \frac{a}{f_a}
    \frac{\alpha_{s}}{8\pi}
    G^a_{\QCD, \mu\nu}\widetilde{G}^{a,\mu\nu}_{\QCD}
    +\frac14 g_{a\gamma\gamma}
     a F_{\EM,\mu\nu}\widetilde{F}_{\EM}^{\mu\nu} \, ,
    %+ c_{h}G^\prime\tilde{G}^\prime+...\right] \, .
    \label{eq:aGGcouplings2}
\end{align}
where
\begin{align}
    g_{a\gamma\gamma}
    &=
    \frac{\alpha_{\EM}}{2\pi f_a}
    \left(\frac{E}{N}-1.92\right).
    \label{eq:QCDaxionphoton}
\end{align}
For the simplest embedding -- where the SM gauge group is contained in an $SU(5)$ subgroup of $\GUT$ -- the ratio is fixed to be $E/N = 8/3$ (see appendix~\ref{sec:embedding} for the more general case).
%\textcolor{red}{Mario: I wonder if this should be moved to Eq. 1.1 -- On the other hand, the shift produced by the numerical factor in (\ref{eq:QCDaxionphoton}) is the usual, calculable IR contribution from the mixing with mesons, which is model independent.} 
Thus simple GUTs make a sharp prediction for the relative couplings of the QCD axion to photons and gluons. The axion-gluon coupling produces a mass for the axion through QCD effects%~\cite{diCortona:2015ldu}
\begin{align}
    m_{\QCD}
    &=
    \frac{ m_\pi f_\pi}{f_a}
    \frac{1}{\sqrt{z}+\frac{1}{\sqrt{z}}} 
    \,,
    \qquad 
    z = (m_u/m_d).
    \label{eq:QCDmass}
\end{align}
%defining the QCD line in the $g_{a\gamma\gamma}-m_\QCD$ space. 
Requiring a solution to the strong $\CP$ problem implies that this dominates the mass of the axion, giving a one-to-one relationship between the axion-photon coupling and the axion mass, defining the QCD line for the the GUT axion. 

The coupling in equation~\eqref{eq:QCDaxionphoton} is a combination of an anomaly coefficient and a calculable component from mixing of the axion with mesons and is independent of whether the $\PQ$ spontaneous symmetry breaking scale is above or below the GUT scale. The $\upq$ current divergence above the scale of $\PQ$ and GUT breaking is
\begin{align}
    \partial^\mu J^{\PQ}_\mu
    &=
    \mathcal{A}\frac{\aGUT}{8\pi}
    \GGUT_{\mu\nu}\widetilde{\GGUT}^{\mu\nu} \, .
\end{align}
In the spontaneously broken phase, the anomalous variation of the effective action under the $\PQ$ symmetry arises from the shift of the axion in the terms in equation~\ref{eq:aGGcouplings}. This is analogous to the WZW term in the QCD chiral Lagrangian that matches the EM anomaly of the neutral component of the chiral current. The only difference between an axion decay constant larger or smaller than the GUT scale is that for $M_{\rm GUT}>f_a$ one matches $\partial^\mu J^{\PQ}_\mu$ directly to the right side of equation~\eqref{eq:aGGcouplings}.

This prediction can be altered by introducing new sources of $\upq$ symmetry breaking. In order to retain the axion solution to the strong $\CP$ problem these contributions to the potential must either be strongly subdominant to the QCD contribution or aligned with the QCD vacuum. QCD generates a potential for the axion of the form
\begin{align}
   V_\QCD(a) \simeq f_\pi^2 m_\pi^2 
   \left(1 - \cos \left(\frac{a}{f_a} + \bar{\theta} \right) \right) \, .
   \label{eq:V_QCD}
\end{align}
This potential explicitly breaks the shift symmetry of the axion and dynamically sets the effective $\theta$-angle, defined as
\begin{align}
   \theta_{\eff} &\equiv \frac{ \langle a \rangle }{f_a} + \bar\theta \, ,
   \label{eq:thetaeff}
\end{align}
to zero. If there are additional terms which break the $U(1)_\PQ$ symmetry they will generically lead to $\theta_{\eff} \neq 0$, and therefore must be strongly suppressed relative to the QCD contribution to the potential in order to satisfy the severe constraints from experimental searches of neutron EDMs~\cite{Pendlebury:2015lrz}:
\begin{align}
    \theta_\eff \lesssim 10^{-10} \, .
    \label{eq:thetabound}
\end{align}

\subsection{Additional Instantons}
\label{subsec:quality}

One way additional sources of PQ breaking can arise is if there are additional confining gauge groups $G'$ embedded in $\GUT$. The QCD axion will couple to the $G'$ gauge group through the term
\begin{align}
   \mathcal{L}
   \supset 
   N_{h}\frac{a}{F_a} G^\prime\widetilde{G}^\prime
   \, ,
\end{align}
where $N_h$ is an anomaly coefficient. After confinement, $G'$ instantons will generate a potential for the axion of the form which will generically be offset by an angle $\delta$ from the QCD vacuum\footnote{Similar potentials are also generated if $G^\prime$ is spontaneously broken at a high scale \cite{Agrawal:2017ksf} with a possible chiral and exponential suppression due to a weak gauge coupling.}:
\begin{equation}
\begin{aligned}
   \Delta V(a) &= \frac{N}{N_h}{\Lambda'}^4\left(1 -  \cos \left( N_h\frac{a}{F_a} + \theta_h \right) \right) \, ,          \\
   \delta & = \frac{N}{N_h}\theta_h -  \bar \theta \, ,
\end{aligned}
    \label{eq:V_PQquality}
\end{equation}
%which are generated by $SU(N)$ instantons. This could occur if $\GUT$ breaks to $G' \times SU(N)$ with the SM gauge group a subgroup of $G'$ and $SU(N)$ confines at a scale $\Lambda' < M_{\rm GUT} $. This term in the potential has a minimum which is offset from the QCD minimum by an angle $\delta$. 
where the factors of $N, N_h$ in the potential and definition of $\delta$ are included for convenience. The bounds on $\theta_\eff$ require that
\begin{align}
 %   \frac{N_h/N{\Lambda'}^4 \sin(N_h\delta/N)}{\textcolor{red}{ m_\pi^2f_\pi^2} 
 %   +N_h/N {\Lambda'}^4 \cos(N_h\delta/N)} \lesssim 10^{-10},
    \frac{{\Lambda'}^4 \sin(\delta')}{ m_\pi^2f_\pi^2
    +\frac{N_h}{N} {\Lambda'}^4 \cos(\delta')} \lesssim 10^{-10},
    \label{eq:quality}
\end{align}
where $\delta' = N_h\delta /N$, requiring any contributions to the axion potential that are not aligned with the QCD vacuum to be exceptionally small relative to the QCD terms. Assuming a misalignment angle $\delta \sim \mathcal{O}(1)$ the quality problem can be restated in terms of the contributions to the axion mass as:
\begin{align}
    \frac{{m'}^2}{m_\QCD^2} \lesssim 10^{-10} \, .
    \label{eq:quality2}
\end{align}
If $\delta$ is sufficiently small then these additional contributions can be consistent with $\theta_\eff\simeq 0$ while giving a large mass to the QCD axion. This has the effect of moving the QCD axion off the QCD line towards the right while preserving the ratio of couplings to gluons and photons. The misalignment angle $\delta$ vanishes above the GUT scale due to the unification of the SM with $G'$, but can run and become $\mathcal{O}(1)$ at low energies depending on the matter content of the model.

\subsection{Perturbative contributions}

One can also consider the effect of explicit PQ-breaking operators \cite{Kamionkowski:1992mf}
\begin{equation}
    \Delta V_n (a) \sim c_n\frac{F_a^n}{M_P^{n-4}} e^{ia/F_a} +{\rm h.c.}\, ,
    \label{eq:HOops}
\end{equation}
These operators generate a mass for the axion of order
\begin{align}
    {m'}^2 \sim c_n\frac{F_a^{n-2}}{M_P^{n-4}} \, .
\end{align}
If the potential terms~\eqref{eq:HOops} have a minimum which is misaligned with respect to the QCD vacuum by an angle $\delta \sim 1$ then the mass generated by these operators must satisfy the bound~\eqref{eq:quality2}, meaning these contributions must be forbidden up until large $n$. One way to forbid the lower $n$ operators is to impose a $\mathbb{Z}_N$ symmetry under which $\Phi$ is charged non-trivially~\cite{Krauss:1988zc,Kamionkowski:1992mf}. This forbids all operators~\eqref{eq:HOops} for $n \leq N$ and therefore improves the `quality' of the global $\PQ$ symmetry. In any case, we again retain the prediction of the ratio of gluon-photon couplings.

$\PQ$-violating operators may also alter the predictions above by shifting the ratio $E/N$. At dimension 6 there is an operator of the form:
%Already at dimension 6, one could try with an operator of the form:
%
\begin{equation}
    \frac{c}{\Lambda} e^{ia/F_a}
    \text{Tr}[\Sigma \, \GGUT\widetilde{\GGUT}] + \text{h.c.}
    = \frac{\text{Im}(c)}{\Lambda}  \frac{a}{F_a}
    \text{Tr}[\Sigma \, \GGUT\widetilde{\GGUT}] + \ldots \,,
    \label{eq:operator2}
\end{equation}
where the scalar $\Sigma$ is an adjoint field taking a non-zero VEV in the hypercharge direction. We expect that these operators are suppressed by $\Lambda$ at the Planck scale or perhaps below if the fundamental scale lies between $M_{\rm GUT}$ and the $M_P$ (see for example~\cite{Kolda:1996ea}). As a concrete example, for $SU(5)$ the adjoint VEV reads
\begin{equation}
    \langle\Sigma\rangle \propto \text{diag}(-1/3,-1/3,-1/3,1/2,1/2)\,.
\end{equation}
This operator induces a shift to the previously obtained predictions by inducing an axion coupling to photons of size $g_{a\gamma\gamma}\sim c\frac{\langle{\Sigma}\rangle}{\Lambda}\frac{\alpha_{\EM}}{f_a}$. 
%Assuming  $c\sim O(1)$ and a cut-off at around the Planck scale $\frac{\langle{\Sigma}\rangle}{\Lambda}\sim 10^{-2}$ the departures would lie at the percent level. 
For axions that do not couple to GUT bosons through an anomaly, this could lead to an observable photon coupling. However, this operator necessarily breaks the axion shift-symmetry and induces a perturbative mass for the axion which can be estimated as:
\begin{equation}
    m^2\sim c \alpha_{\rm GUT}^2 \Lambda^2 \, .
\end{equation}
%For the ALP to be light $c$ must be very small, \textcolor{red}{calling for a non-perturbative origin of the operator with the associated instanton-like exponential suppression,} severely suppressing the contribution to the axion-photon coupling.
For the ALP to be light $c$ must be very small,  severely suppressing the contribution to the axion-photon coupling. This could happen naturally if the operator in equation~\eqref{eq:operator2} is generated by non-perturbative effects and comes with an exponential suppression factor.

%\MN{I would say something more like: For the ALP to be light $c$ must be very small,  severely suppressing the contribution to the axion-photon coupling. This could happen naturally if the operator in equation~\eqref{eq:operator2} is generated by non-perturbative effects and comes with an exponential suppression factor.}

\subsection{Discrete symmetries, mirror worlds and clockwork}

\label{sec:discrete}

The strong-$\CP$ problem may also be solved using discrete symmetries. The simplest example of this is the Nelson-Barr mechanism~\cite{Nelson:1983zb, Barr:1984qx} using $\P$ and $\CP$ discrete symmetries. In this case, the axion can be heavy and decoupled without affecting the solution to the strong-$\CP$ problem.

In cases where the axion and a discrete symmetry together solve the strong-$\CP$ problem, the axion mass may be larger or smaller than that predicted by QCD. In~\cite{Hook:2018jle,DiLuzio:2021pxd} a $\mathbb{Z}_N$ symmetry relating different copies of the SM (mirror worlds) was shown to lead to a light QCD axion. In these models the axion mass scales as $\frac{m_a}{m_{\QCD}}\sim \frac{1}{2^{N/2}}$, with $N$ the number of SM copies or GUT-like sectors. The gauge group in these models is
\begin{align}\label{eq:group_mirror_worlds}
    \GUT_1 \times \ldots \times \GUT_N
\end{align}
where the SM is embedded into one $\GUT$ factor. A QCD axion which is much lighter than $m_\QCD$ then requires a large number of SM copies and these copies must not be unified with the SM\footnote{Further unification of the different $\GUT_i$ factors in (\ref{eq:group_mirror_worlds}) into a simple group reveals the fact that an axion transforming non-trivially under the $\mathbb{Z}_N$ necessarily carries GUT charge. Such possibility leads to a (potentially large) perturbative mass for the axion, as will be shown in Sec.\ref{subsec:emergent}.}. The contribution to the axion potential from new, UV instantons of the mirror world can also be aligned respect to QCD by using discrete symmetries. This has led to different models~\cite{Rubakov:1997vp,Berezhiani:2000gh,Hook:2014cda, Hook:2019qoh} where the axion is substantially heavier than usual while still solving the strong CP problem. 

%The possibility of a very light QCD axion has been recently explored in~\cite{Hook:2018jle,DiLuzio:2021pxd} by using a discrete symmetry, $\mathbb{Z}_N$, relating different copies of the SM (mirror worlds). In these models the axion mass scales as $\frac{m_a}{m_{\QCD}}\sim \frac{1}{2^{N/2}}$, with $N$ the number of SM copies - or GUT-like sectors if we assume an underlying GUT in the UV. While it is not logically impossible, getting a light QCD axion in this way requires a large extension of the GUT philosophy.

%If the UV gauge groups is unified into a simple group, we (see also an example

%\subsection{Mirror worlds and $\mathbb{Z}_N$}

%\subsection{Clockwork axions}

A different mechanism that has been proposed to increase the QCD axion coupling to photons, mimicking an ALP, is the clockwork mechanism~\cite{Choi:2015fiu,Kaplan:2015fuy,Farina:2016tgd,Agrawal:2017cmd}. This mechanism works by introducing $n+1$ complex scalars $\phi_i$ and $n+1$ U(1) symmetries, with scalar $\phi_i$ charged under the symmetries $(U(1)_i,U(1)_{i+1})$ with charge $(1,q)$. The scalars at either end of the chain, $\phi_0$ and $\phi_n$, couple to fermions charged under different gauge groups, e.g.~QCD and EM respectively. The $U(1)$ symmetries are broken to a diagonal $\upq$ symmetry under which the $n^{\rm th}$ scalar has $\PQ$ charge $q^{n}$. Since the fermions which generate the electromagnetic anomaly couple only to $\phi_n$ and the fermions which generate the QCD anomaly couple to $\phi_1$ then the ratio of anomalies scales like 
\begin{align}
    \frac{E}{N} \propto q^{n} \, ,
\end{align}
leading to an exponential hierarchy in the anomaly coefficients. This enhancement clearly relies on coupling the fermions that generate the electromagnetic and color anomalies to different scalars, which is not possible if the SM is unified and the fermions form complete GUT multiplets.
%If one works with full $\GUT$ multiplets, for example, no enhancement is possible and one gets the usual prediction for a standard embedding $E/N=8/3$.

\subsection{Measuring the QCD axion couplings}

A measurement of $g_{a\gamma\gamma}$ for the QCD axion will tell us the value of the ratio $E/N$, so can provide an indirect probe of grand unification. As above, for a level one embedding GUT theories predict $E/N = 8/3$. Extracting this ratio will be the most immediate target following any discovery. Measuring this ratio precisely is experimentally challenging, but a measured value far from the GUT prediction will be strongly at odds with unification, even with a large experimental uncertainty.

The measurement of this coupling is an interesting experimental question. If the axion decay constant is $f_a \simeq 10^{12}\GeV$, then the axion lies in the band of currently operating cavity resonance haloscopes~\cite{ADMX:2018gho, ADMX:2019uok, ADMX:2021nhd}. The sensitivity of axion haloscopes  is set by the limited time that the resonant cavity is tuned to a specific mass. Upon discovery, a large amount of data at the axion mass can be gathered, which can measure the rate of axion-photon conversion to a high precision. The rate is proportional to,
\begin{align}
    P_{a\to\gamma}
    &\propto
    \rho_{\rm a} g_{a\gamma\gamma}^2
    \,,
\end{align}
where $\rho_a$ is the local axion density~\cite{Sikivie:1983ip, Sikivie:1985yu}. The narrow bandwidth of the cavity means that the resonance condition corresponds to a precise measurement of the axion mass. Thus, haloscopes are in a position to measure the ratio $g_{a\gamma\gamma}/m_a$ to test equation~\ref{eq:QCDline1}. Unfortunately the quantity $\rho_a$ is difficult to estimate even if we assume the axion makes up all of dark matter (DM), as the local DM density is not known precisely~\cite{Frandsen:2011gi,Read:2014qva,Green:2017odb,2022MNRAS.511.1977S, Buch:2018qdr,2020MNRAS.495.4828G, 2020A&A...643A..75S,2021A&A...653A..86W}. This presents a challenge to using QCD axion couplings in this way as a precision test of GUTs. It will be interesting to consider the possibility of identifying a sub-component of DM  with a more accurate local density prediction, such as axions trapped within the Earth's or Sun's gravitational basin~\cite{VanTilburg:2020jvl,DeRocco:2022jyq}. Even if it is a subdominant component, it is possible that it can be differentiated from the halo dark matter through velocity measurements~\cite{Millar:2017eoc}, and that the haloscope sensitivity may be enough to extract a measurement of $g_{a\gamma\gamma}$.

Another class of proposed haloscope experiments look for axion DM through its coupling to gluons to measure a time-dependent oscillating electric dipole moment (EDM). These experiments are a way to measure the axion abundance, but are limited by theoretical uncertainties on nuclear matrix elements which are only known to about 30\%~\cite{Pospelov:1999mv}. Another potential challenge, depending on the value of $f_a$, is that the EDM experiments and the photon experiments may not both be sensitive to the QCD axion however a region of overlap does exist. If the axion exists in this overlap region more precise calculations of the matrix elements can break the degeneracy between the couplings and the local DM density. Other observables which use derivative axion couplings do not help since these couplings are not quantized and undergo renormalization.

We see that GUTs provide a very precise target for the QCD axion, but utilizing the relation between the photon and gluon coupling to test aspects of unification is experimentally challenging. It will be interesting to develop other strategies that can measure and test the quantized couplings after an initial discovery. %While measuring consequences of the integer $\mathcal{A}$ for the QCD axion is difficult, it turns out that measuring ``0" for the other axion-like particles can be more accessible experimentally. This is the case we now describe.

\section{GUT prediction for axion-like particles}

\label{sec:GUTALP}

In this section we consider the possibilities for generating an axion-photon coupling for axions which have vanishing anomaly with $\GUT$. This can occur through mass or kinetic mixing of axions, axions charged under $\GUT$ or mixing between GUT gauge bosons and a hidden gauge sector. For the case of axion mixing we find that while it is possible to have a light ALP, the ALP coupling to photons is suppressed proportional to the squared ALP mass in the low-mass limit. Charged ALPs receive contributions to their mass from integrating out heavy gauge bosons or additional instantons and generically must be heavier than $m \gtrsim 1 \GeV$ to satisfy current experimental constraints. Kinetic mixing of gauge bosons generates an ALP-photon coupling which is proportional to the (small) mixing parameter $\epsilon^2 \lesssim 10^{-8}$.

Unification of the SM in the UV therefore rules out the possibility of very light ALPs with experimentally accessible couplings to photons in the absence of tuning. This points to a preferred region of parameter space at masses greater than the QCD axion expectation for ALP searches which rely on the photon-coupling. ALP couplings to fermions are unsuppressed, however, so experiments~\cite{Budker:2013hfa,Arvanitaki:2014dfa} which probe ALP couplings to SM fermions offer a promising avenue to search for light ALPs in GUT theories. In this way, GUT theories can naturally lead to an ALP phenomenology similar to the case of the photophobic ALP~\cite{Marques-Tavares:2018cwm,Craig:2018kne}.

\subsection{ALPs through the QCD axion portal}

\label{subsec:mixing}

If the $\upq$ symmetry\footnote{By $\PQ$ symmetry we refer to the $U(1)$ which is anomalous with $\GUT$. In equation~\eqref{eq:Naxion} the axion under this symmetry is the combination $ \sum_{i=1}^N \frac{a_i}{f_i}$.} mixes with other global $U(1)$ symmetries then it is possible for some axions to couple to photons without obeying equation~\eqref{eq:QCDline1}. This manifests itself as mass mixing in the axion potential, which for $N$ axions $a_i$ is given by
\begin{align}
    V(a_i) &= \left( \sum_{i=1}^N \frac{a_i}{f_i} \right) 
    \GGUT \tilde \GGUT
    + \frac{1}{2} M^2_{ij} a_i a_j \, .
    \label{eq:Naxion}
\end{align}
The axions may also mix kinetically, but after redefining fields to move to a basis where the axion kinetic terms are canonical the effects of kinetic mixing are included in the potential~\eqref{eq:Naxion} if we allow arbitrarily large decay constants~\cite{Agrawal:2018mkd, Fraser:2019ojt}. In order to not introduce a quality problem there must be a $U(1)$ subgroup of $\upq \times \prod_i U(1)_i$ that remains unbroken except due to QCD effects, which is true if the $M^2$ matrix has a zero eigenvalue.

%nother possible way in which an ALP may couple to photons without a mass which falls on the QCD line could be due to mixing of $\upq$ with other global $U(1)$ symmetries, which may be explicitly broken. The most general (\MN{\tt{maybe not true?}}) case where this occurs is for a set of axions $a_i$ labelled by $i\in [1, N]$, each of which couples to the full GUT gauge group with independent decay constants and additional mass terms. The potential which describes this is: 
%\begin{align}
%    V(a_i) &= \left( \sum_{i=1}^N \frac{a_i}{f_i} \right) G_\GUT \tilde G_\GUT 
%    + \frac{1}{2} M^2_{ij} a_i a_j \, ,
%\end{align}
%after redefining the $f_i$ to absorb the anomaly coefficients $\mathcal{A}_i$. In order to avoiding introducing a quality problem the $M^2$ matrix must have a zero eigenvalue. This guarantees there is an exact $U(1)_{\PQ}$ symmetry which is unbroken except by QCD instantons, this solving the strong CP problem. 

An illustrative example is the case of two mixed axions with potential
\begin{align}
    V(a_1, a_2) &= \left(\frac{a_1}{f_1} + \frac{a_2}{f_2} \right) 
    \GGUT \tilde \GGUT
    + \frac{1}{2} m_2^2 a_2^2 \, .
    \label{eq:two_axion_mixing}
\end{align}
After QCD confines the mass terms in the potential are:
\begin{align}
    V_{\rm mass} = \frac{f_\pi^2 m_\pi^2 }{2} \left( \frac{a_1}{f_1} + \frac{a_2}{f_2} \right)^2
    + \frac{1}{2} m_2^2 a_2^2 \, ,
\end{align}
and the coupling to QCD induces mass mixing between $a_1$ and $a_2$. In this case we find that there is always a QCD-like axion $a$ with a mass and photon coupling set by $f_\eff$, where
\begin{align}
    \frac{1}{f_\eff^2} = \frac{1}{f_1^2 }+ \frac{1}{f_2^2 } ,
    \label{eq:f_eff}
\end{align}
and a second mass eigenstate $b$ which may be lighter or heavier depending on the value of $m_2^2$.

In the limit $m_2 \ll f_\pi m_\pi / \max(f_1, f_2)$, the QCD axion is identified with
\begin{equation}
    \frac{a}{f_\eff} \simeq \frac{a_1}{f_1} + \frac{a_2}{f_2}
\end{equation}
and the orthogonal field $b$ is approximately massless with negligible coupling to photons. More precisely, in this limit the coupling of the light axion to photons is suppressed relative to the QCD axion by a factor 
\begin{equation}
    g_{b \gamma \gamma} \propto \frac{m_{b}^2\times \max(f_1^2, f_2^2)}{f_\pi^2 m_\pi^2}.
\end{equation}
In the opposite limit, when $m_2 \gg f_\pi m_\pi / \min(f_1, f_2)$, then the QCD axion is identified with $a \simeq a_1$ and the second axion $b \simeq a_2$ couples to photons with a coupling of similar magnitude but is much heavier than the QCD axion. When $m_2^2 \simeq m_\QCD^2$ both mass eigenstates have a mass and photon coupling which falls close to the QCD line.

%For light axions, however, the induced coupling to photons is suppressed by the factor $m_{\rm light}^2/m_\QCD^2$, where $m_{\rm light}$ is the low mass eigenstate. Ultralight axions therefore have strongly suppressed couplings to photons, so the deviation from the expected mass-coupling relationship is at most $\mathcal{O}(1)$ for any axions which may be observable in upcoming experiments. (\MN{\tt{may need to check/qualify this statement.}})

\begin{figure*}[t]
	\centering
	\includegraphics[width=0.85\textwidth]{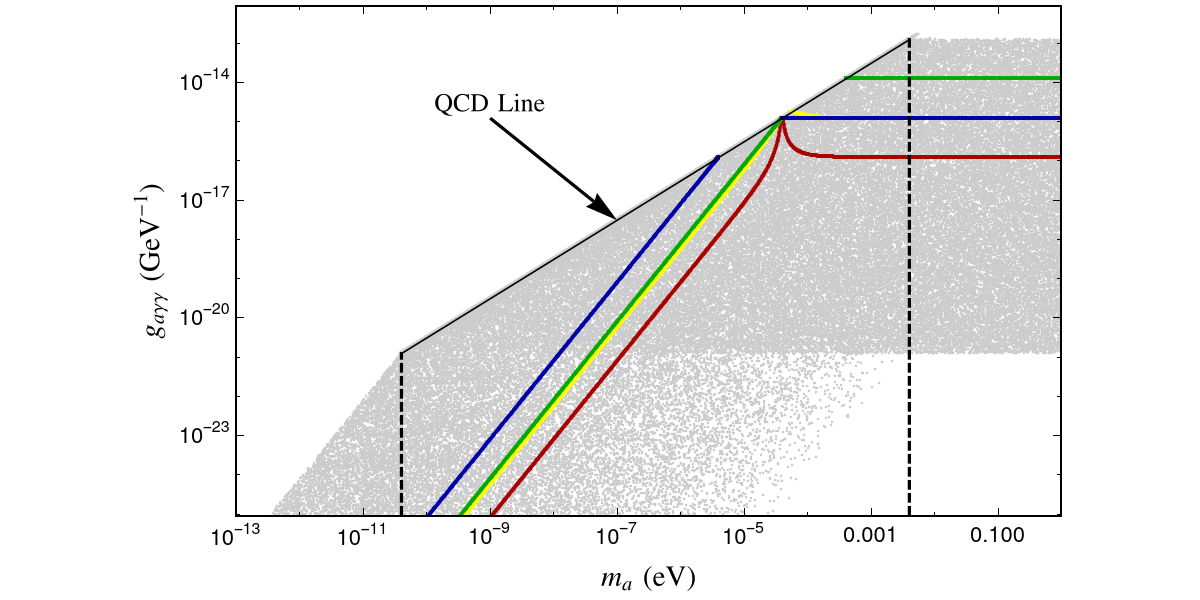}
%	\includegraphics[width=0.6\textwidth]{Figures/mixing2.pdf}
%	\caption{Figure showing the mass and photon coupling of the non-QCD mass eigenstate in the two axion model of equation~\eqref{eq:two_axion_mixing}. The decay constants are fixed to be $f_1 = f_2 =10^{12} \GeV$ and points are generated by varying the parameter $m_2$.  \MN{\tt{Numbers are not quite correct}}}
    \caption{Scatter plot showing the mass and photon coupling of the two mass eigenstates in the model described by the potential~\eqref{eq:two_axion_mixing}. Shown in light grey are the points generated, the dashed lines show the maximum and minimum masses for the QCD-like eigenstate and are set by the chosen range of $f_1, f_2$. The thin black line shows the QCD axion line. The coloured lines show the points generated for fixed decay constants: the red line shows the case where $f_1 = 10^{12} \GeV, f_2 = 10^{13} \GeV$, the yellow line shows $f_1 = f_2 = 10^{12} \GeV$, the green line shows $f_1 = 10^{12} \GeV, f_2 = 10^{11} \GeV$ and the blue shows $f_1 = 10^{13} \GeV, f_2 = 10^{12} \GeV$.}
	\label{fig:mixingplot}
\end{figure*}

The photon coupling and mass of the two mass eigenstates are shown in figure~\ref{fig:mixingplot} for multiple combinations of the parameters $f_1, f_2, m_2$. The grey points are generated by taking randomly generated values of the parameters $f_1, f_2$ logarithmically spaced in the range $[10^{10}, 10^{18}] \GeV$ and $m_2$ logarithmically spaced in the range $[10^{-11}, 1] \eV$. Allowing for larger (smaller) values of $m_2$ simply increases the number of points away from the QCD line to the right (bottom) of the plot.

The coloured lines in figure~\ref{fig:mixingplot} are generated by fixing $f_1$ and $f_2$ and varying $m_2$ in the range $[10^{-11}, 1] \eV$. The yellow line shows points where $f_1 = f_2 = 10^{12}\GeV$. In this case, there is always a QCD axion with mass given by equation~\eqref{eq:QCDmass} with $f_a=f_
\eff$ and a second eigenstate with the same coupling but larger mass (the horizontal line to the right of the point on the QCD line) or with coupling suppressed by the factor $m^2_b/m_\QCD^2$ (the line approaching the QCD line from below). The other coloured lines each show various cases for fixed $f_1 \neq f_2$ with $m_2$ scanned. In each case there is always a QCD axion with coupling and mass set by $f_\eff$.

These results clearly point to an allowed range of parameter space for ALPs at masses heavier than the QCD axion (to the right of the QCD line). A light ALP with sizeable coupling to photons is not generated via mass mixing. This can be understood as follows: in order to have a sizeable coupling to photons the non-QCD axion must contain a sizeable admixture of the combination $\frac{a_1}{f_1} + \frac{a_2}{f_2}$, giving a contribution to $m_b^2$ of order $m_\QCD^2$.

The points to the right of the QCD axion, corresponding to the points where $m_2$ is large, form a flat line with coupling to photons given by:
\begin{equation}
    g_{b\gamma \gamma} =
    \frac{\alpha_{\EM}}{2\pi f_2}
    \left(\frac{E}{N}-1.92\right),
\end{equation}
and correspond to the case where $a_2 $ is heavy and $a_1$ is identified with the QCD axion to a good approximation. The points below the QCD line correspond to the cases where $m_2$ is small and form a line described by
\begin{equation}
    g_{b \gamma \gamma} \simeq g_{a \gamma \gamma} 
    \frac{m_{b}^2\times \max(f_1^2, f_2^2)}{f_\pi^2 m_\pi^2} \, ,
\end{equation}
approaching the QCD line at a point where the coupling and mass are set by $\max(f_1, f_2)$. When $f_1>f_2$ (blue and green points) it is possible that there are two axions on the QCD line with different masses if $m_2$ is tuned appropriately.

A similar conclusion holds for multiple axions coupling to gauge bosons through the QCD axion portal. For the general case of $N$ axions the mass terms in the potential can be written as
\begin{align}
    V_{\rm mass}= \frac{f_\pi^2 m_\pi^2}{2} \left(\frac{a_0^2}{f_0^2}\right)
    + \frac{1}{2}  a^T_j M^2_{ij} a_i \, ,
    \label{eq:N_axion_mixing}
\end{align}
for $i, j \in 0, \ldots N-1$, where $a_0$ is the axion of the anomalous $\PQ$ symmetry. In this basis, the only axion which couples to photons is $a_0$ with coupling constant
\begin{align}
    g_{a \gamma \gamma } &=
    \frac{\alpha_{\EM}}{2\pi f_0}
    \left(\frac{E}{N}-1.92\right)  \, .
\end{align}
We can then rotate to the basis of mass eigenstates $b_i$ with an orthogonal matrix $R$:
\begin{align}
    a_i = R_{ij} b_j \, .
\end{align}
The mass and coupling to photons of the $i$'th eigenstate is
\begin{align}
    &m_i^2 = \frac{f_\pi^2 m_\pi^2}{ f_0^2} R_{0i}^2
    + R^T_{ij} M^2_{jk} R_{ki} \, ,
    &&g_{b_i \gamma \gamma } =
    \frac{\alpha_{\EM}}{2\pi f_0}
    \left(\frac{E}{N}-1.92\right) R_{0i} \, ,
\end{align}
where there is no sum over $i$ in the first term. This basis makes it clear that the QCD line provides an upper bound on the ALP parameter space in GUTs.

The results from the two-axion case continue to apply qualitatively in the general case as well. This is evident in the $a$ basis where we further rotate $a_i$ for $i \geq 1$ to diagonalise the components of $M^2_{ij}$ with $i \geq 1, j \geq 1$, leaving only pairwise mixing between $a_0$ and the orthogonal states. As in the two axion case if the mixing term is large the corresponding ALP can be integrated out and corresponds to the heavy ALPs on the horizontal lines shown in figure~\ref{fig:mixingplot}. In the light mass regime the ALP will again have a coupling to photons suppressed by $m_{\rm light}^2/m_\QCD^2$.

\subsection{GUT charged axions}

\label{subsec:emergent}

%\MN{\tt{I'm thinking this might deserve it's own section. It doesn't quite fit in with mass mixing etc, as it dramatically changes the prediction for the mass vs. coupling rather than introducing a deviation around it, is also not quite a GUT group because of the confining hidden sector.}}

The discussion of section~\ref{sec:GUTQCDaxion} presumed that the axion was a GUT singlet and therefore coupled to $\GUT$ only through the anomaly coefficient. In the low energy theory there may be additional pseudo Nambu-Goldstone bosons (pNGBs) which are singlets under the SM but not under $\GUT$. The simplest possibility is an elementary scalar in the adjoint representation of $\GUT$. However, in this case GUT gauge boson loops will generate a perturbative mass proportional to the UV cutoff. Supersymmetry may reduce this mass further, but scalar masses parametrically below $m_{\rm susy}$ seem to be unlikely.

A potentially interesting ALP candidate with mass below the UV cutoff can occur in composite axion models~\cite{Kim:1984pt,Choi:1985cb,Kaplan:1985dv,Randall:1992ut}. Just as pions emerge as pNGBs after QCD confinement breaks the flavour symmetry of the quark sector, the axion can emerge as the pNGB of a broken flavour symmetry of the hidden sector. After the spontaneous breaking of the GUT symmetry this axion gets an irreducible mass from integrating out the heavy GUT gauge bosons. In this case the axion can couple to a different linear combination of the SM gauge groups than the QCD axion. Its mass is protected by compositeness but there remains an irreducible contribution to the axion mass from gauge boson loops which rules out a light axion. It is also possible to include elementary axions in addition to the composite axion(s), but the phenomenology of the singlet states is not changed from the discussion in previous sections.

We now describe an example model which illustrates the physics of the composite axion. The gauge group we consider is
\begin{equation}
    \GUT \times SU(N) \, ,
    \label{eq:G_emergent}
\end{equation}
with $\GUT$ a simple group containing the SM. We take all the SM matter fields to be singlets under $SU(N)$, with $SU(N)$ confining at a scale scale $\Lambda_{N}< M_{\rm GUT}$\footnote{Note that if $\GUT$ and $SU(N)$ are unified at a UV scale, then instantons of the new sector may reintroduce the PQ quality problem (see section \ref{subsec:quality}).}. The additional matter content of the model consists of fermion fields with charges under
$(\GUT, SU(N))$ given by:
\begin{align}
    &\Psi \sim  (\Box,N), 
    &&\bar \Psi \sim (\bar{\Box},\bar{N}) && \psi\sim (1,N) && \bar \psi\sim (1,\bar{N}) \,,
\end{align}
where the $\Box$ indicates the fundamental representation of $\GUT$. Taking $\GUT = SU(5)$ as the canonical example, there is a flavour symmetry
\begin{align}
    U(6)_L \times U(6)_R
    = SU(6)_L \times SU(6)_R \times U(1)_V \times U(1)_A\,.
\end{align}
$U(1)_V$ is exact and is the equivalent of baryon number under the hidden sector gauge group and $U(1)_A$ is explicitly broken by $SU(N)$ instantons. After confinement the fermions condense and get an expectation value
\begin{equation}
    \left \langle \Psi\bar{\Psi} \right \rangle=\left \langle \psi\bar{\psi} \right \rangle
    =\Lambda_N^3  \,,
    \label{eq:Emergentvev}
\end{equation}
spontaneously breaking the flavor symmetry down to
\begin{align}
     SU(6)_L \times SU(6)_R \times U(1)_V \times U(1)_A\,
     \to SU(6)_{V}\times U(1)_V .
\end{align}
The $SU(6)_V$ flavour symmetry is explicitly broken by the weak gauging of the GUT group. The pNGBs arising after spontaneous breaking of the flavor symmetry transform in the adjoint representation of $SU(6)_V$, $\mathbf{35}$, and decompose into fields with  GUT $SU(5)$ charges:
\begin{equation}
    \mathbf{35}\rightarrow \mathbf{24}+\mathbf{5}+\mathbf{\bar{5}}+\mathbf{1} \, .
\end{equation}
The GUT singlet remains massless perturbatively and behaves as a QCD axion, coupling diagonally to $\GUT$. The pNGBs in the fundamental, $\mathbf{5}+\mathbf{\bar{5}}$, all have SM charges and get mass contributions from SM gauge boson loops of order:
\begin{equation}
    m \sim g_{\rm sm} \Lambda_N \, ,
    \label{eq:pNGB_mass}
\end{equation}
where $g_{\rm sm}$ is a SM coupling\footnote{The coupling, or combination of couplings, appearing in equation~\eqref{eq:pNGB_mass} depends on the charges of the pNGB under the SM gauge group.}~\cite{Choi:1985cb} (see \cite{Dimopoulos:2016lvn} for a recent discussion).

The pNGB transforming as a GUT adjoint, $\mathbf{24}$, which we call $\varphi^a$, splits under the SM gauge group as:
\begin{equation}
    \varphi^a 
    \sim (\mathbf 8,\mathbf 1)_0+(\mathbf 1,\mathbf 3)_0+(\mathbf 3, \mathbf 2)_{5/6}+(\bar{\mathbf{ 3}},\mathbf 2)_{-5/6}+ (\mathbf 1,\mathbf 1)_0\,.
\end{equation}
As for the $\mathbf{5}+\mathbf{\bar{5}}$ above, those states with SM charge get a large mass induced by SM gauge boson loops, as in equation~\eqref{eq:pNGB_mass}. The SM singlet $\varphi^{24}$, on the other hand, is identified as a composite axion and appears massless at this level. However, after integrating out the heavy GUT gauge bosons our EFT contains a 4-fermion operator:
\begin{align}
    \frac{\alpha_{\rm GUT}}{M^2_{\rm GUT}} 
    \left( \Psi \bar \Psi \right)
    \left( \Psi \bar \Psi \right).
\end{align}
Expanding around the vev~\eqref{eq:Emergentvev}
%begin{align}
%    \left \langle \psi\bar{\psi} \right \rangle
%    =\Lambda_N^3 \left( 1+ \frac{\varphi^a T^a}{\Lambda_N} \right) \,,
%\end{align}
leads to mass terms\footnote{One could ask if SUSY can further protect the mass of the ALP. However, in this kind of scenarios SUSY may be dynamically broken by fermion condensates of the new confining interaction~\cite{Dine:1981za}. Indeed, as shown in~\cite{Affleck:1984xz}, it is generically expected that SUSY is dynamically broken unless there are flat directions for the potential. The study of specific models is interesting by itself and will not be pursued here.} for $\varphi^{24}$ which scale as
\begin{equation}
    m_{\varphi^{24}}\sim
    \alpha_{\rm GUT}\left(\frac{f_a}{10^{10} \GeV}\right)^2
    \left(\frac{10^{16}\GeV}{M_{\rm GUT} }\right)
    10 \, \, \TeV\,,
    \label{eq:Compositemass}
\end{equation}
where the decay constant for the composite axion is $f_a \sim \Lambda_N$. This result coincides with the estimate of~\cite{Vecchi:2021shj,Contino:2021ayn}, where a similar flavor symmetry appears in a different context.

\begin{figure*}[t]
	\centering
	\includegraphics[width=0.7\textwidth]{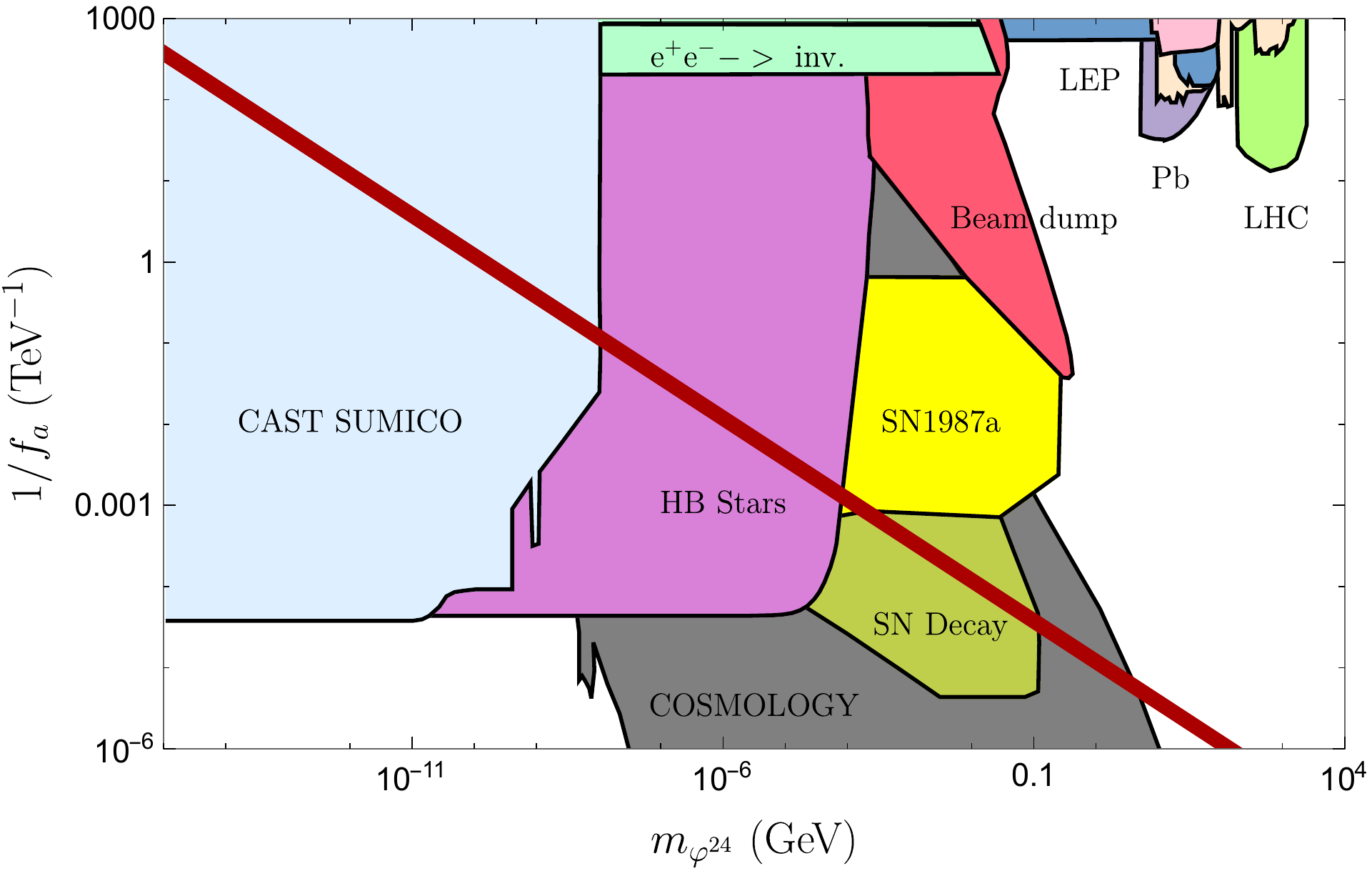}

	%\caption{\MN{\tt{placeholder, need to show experimental bounds on emergent axion}}}
		\caption{ Figure showing bounds on the ALP parameter space with the prediction of the composite ALP shown as the dark red band. Note that the bound from cosmology (gray color) depends on the emergent ALP being the dominant DM component. If its abundance is negligible, then SN constraints are the strongest, implying $m_{\varphi^{24}}>0.1$ GeV in that case.}

	\label{fig:emergent}
\end{figure*}

The composite axion field $\varphi^{24}$ couples to a different linear combination of the SM gauge bosons than the QCD axion. $\varphi^{24}$ parametrises $SU(5)$ transformations in the hypercharge direction, and the couplings to GUT gauge bosons are set by the anomaly coefficients $\text{Tr}[T^{24}\{T_a,T_b\}]$. In particular the coupling to photons is given by
\begin{align}
    \mathcal{L} = \frac{\varphi^{24}}{F_a} \frac{N \alpha_\EM}{8\pi}
    \frac{4}{9} \sqrt{\frac{3}{5}} F_{\mu\nu} \tilde F^{\mu\nu}\, ,
\end{align}
which comes from the anomaly coefficient with the $W$ bosons of the SM.

The composite axion therefore gives a very different prediction for the relationship between the axion-photon coupling and the axion mass. In particular, $g_{a\gamma \gamma} \propto m_a^{-1/2}$, so lighter axions couple more strongly to photons. As shown in figure~\ref{fig:emergent} much of the low mass parameter space is therefore ruled out, limiting the composite axion mass to be at least of order $m_a \gtrsim 1 \GeV$ with decay constant $f_a \gtrsim 10^5$~TeV.

%\MR{The model with gauge group~\eqref{eq:G_emergent} can also contain a QCD-like axion in addition to the emergent axion. If we add another fermion $\psi$ which is a GUT singlet but charged in the fundamental of $SU(N)$, $\psi\sim (1,N)$, the flavor symmetry breaking pattern reads:
%\begin{equation}
%    SU(6)_L\times SU(6)_R\rightarrow SU(6)_{V}\,,
%\end{equation}
%The pNGBs transform in the adjoint ($\mathbf{35}$) representation of $SU(6)_{V}$ and decompose into fields with charges under the GUT group $SU(5)$
%\begin{equation}
%    \mathbf{35}\rightarrow \mathbf{24}+\mathbf{5}+\mathbf{\bar{5}}+\mathbf{1} \, .
%\end{equation}
%The GUT singlet remains massless perturbatively and behaves as a QCD axion, coupling diagonally to $G'$. }

%\MN{Keep $SU(6)$ as the only example in this section.}

The example above can be generalized for larger flavor symmetries including multiple species in different representations, however, this does not qualitatively change the overall picture. As the flavor symmetry is enlarged multiple copies of GUT-charged and GUT-singlet pNGBs may appear. The charged states get a perturbative mass as detailed above. The singlets couple diagonally to the full $\GGUT$ and follow the discussion in section~\ref{subsec:mixing}.

\subsection{Dark photon - photon mixing }
% In the previous sections we have studied a situation where multiple axions get a mass mixing, and also how charging the axion might change the global picture. We find that these avenues do not provide an arbitrarily light ALP coupled to photons and, instead, that new relations between the ALP mass and the photon coupling appear (see Eqs. (\ref{eq:ALP_mass_coupling}) and (\ref{eq:Compositemass})).

\begin{figure*}[t]
	\centering
	\includegraphics[width=0.4\textwidth]{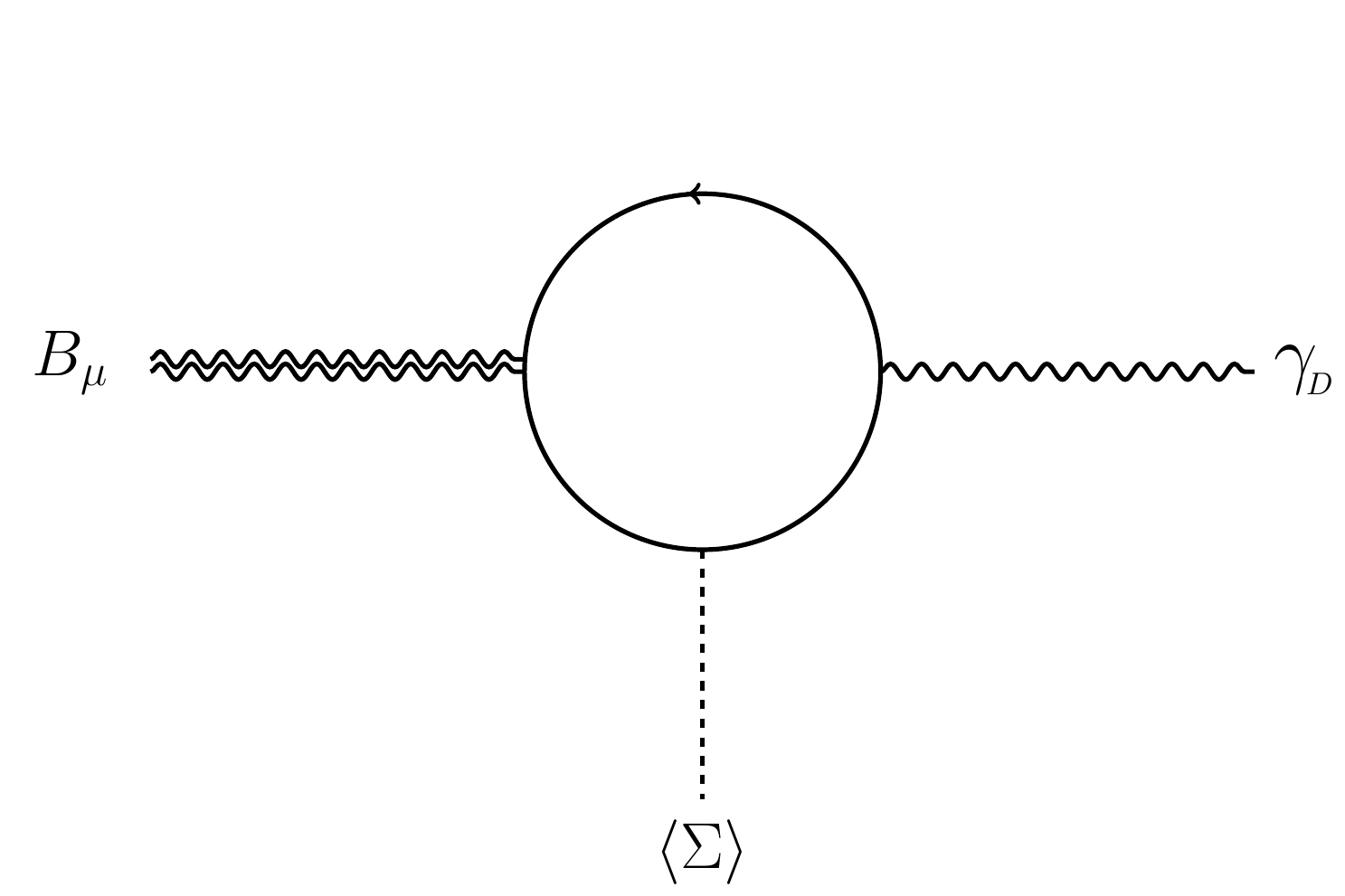}
	\caption{Loop for dark photon-photon mixing. The double wavy line indicates the non-abelian origin of the hypercharge boson. The fermionic line stands for some representation with GUT and dark $U(1)_D$ charge.}
	\label{fig_loop_darkphoton}
\end{figure*}

We consider now the possible effects of photon mixing with a massless dark photon where $U(1)_D$ is not unified with $\GUT$ in the UV\footnote{The situation where $U(1)_D$ is ultimately embedded in the UV GUT group does not modify the result of section \ref{sec:GUTQCDaxion}.}. %(see \cite{Gherghetta:2019coi} for a recent review). 
Assuming that the GUT and dark sector couple to different axions and neglecting axion mixing the axion couplings to gauge bosons are described by the Lagrangian
\begin{equation}
    \mathcal{L}=\alpha_{\rm GUT}\frac{a}{f_a} \GGUT\tilde{\GGUT}
    +\alpha_D\frac{b_{D}}{f_b}F_D\tilde{F_D}\,.
\end{equation}
A tree-level crossed mixing term  between the dark photon and the GUT gauge bosons is forbidden by gauge invariance.  Higher dimensional operators such as: 
\begin{equation}
    O_{\rm mix}= \frac{c}{M_P}F_D \Sigma \GGUT    \,,
\end{equation}
where $\Sigma$ is a scalar field in the adjoint representation, will induce mixing between the dark photon and hypercharge boson after GUT symmetry breaking. For example, when this operator is generated at one loop,  the dimensionless coefficient $c$ is computed by evaluation of the loop diagram in figure~\ref{fig_loop_darkphoton}:
\begin{equation}
c\sim \frac{g_{\rm GUT}g_Dy_\Sigma}{16 \pi^2}  \,,
\end{equation}
and includes the appropriate gauge and Yukawa couplings.  After symmetry breaking, and redefining the photon and dark photon field so that they are canonically normalized, the induced dark axion coupling to photons will be:
\begin{equation}
\frac{\epsilon^2}{8\pi}\alpha_D\frac{b_D}{f_b}F_{\EM}\tilde{F}_{\EM}\,,\,\,\,\text{with: }\,\epsilon^2=\left(\frac{g_{\rm GUT}g_Dy_\Sigma}{16 \pi^2}\frac{\langle\Sigma\rangle}{M_{\rm Pl}}\right)^2    \,.
\end{equation}
Even if we assume a large VEV for the adjoint field, $\langle\Sigma\rangle\sim M_{\rm GUT}$, and order one couplings, $g_{\rm GUT}\sim g_D\sim y_\Sigma\sim O(1)$, we see that the kinetic mixing parameter is approximately given by $\epsilon^2\lesssim 10^{-8}$. The coupling of $b_D$ to photons is suppressed by the small parameter $\epsilon^2$, so deviations from the results of section~\ref{sec:GUTQCDaxion} are small.

%\MN{Are there bounds on $f_b$ in this case?}
%
%\MR{I guess you are worried if we can have small $f_b$ and small alp mass? I wonder if having no symmetry that protects derivative couplings to SM fermions those can be used to constrain the decay constant $f_b$... I think in the photophobia paper they discuss that \cite{Craig:2018kne}. Maybe kinetic mixing also generates these derivative couplings of $b_D$ to fermions. If that is the case, then constraints from fermion couplings apply to $f_b$. If not, and we can go down to $f_b\sim O(1)$ TeV, then we can say that having such ALP correlates with new pheno: dark photon + dark charged matter at such scale.}

Although mixing effects proportional to $\epsilon^2$ are tiny, it might happen that in extreme situations we can get a light ALP to the left of the QCD line. This occurs because when $U(1)_D$ is totally hidden from the visible sector the mass of $b_D$ is a free parameter (although a contribution to the mass may be generated by loops of dark magnetic monopoles~\cite{Fan:2021ntg} or other UV instantons) and the constraints on $f_b$ are weak. The requirements for a light ALP to be generated in this way are a large hierarchy between decay constants, $f_b\sim O(1)\text{ TeV}\ll f_a$, a totally hidden $U(1)_D$ and the absence of kinetic and mass mixing between $a$ and $b_D$. When the latter conditions are not satisfied the $b_D$ ALP may get a coupling to fermions (these couplings are not topological and can be generated by renormalisation effects). Astrophysical probes typically require decay constants $f_b\gg$ TeV for ALPs with couplings to SM fermions~\cite{Irastorza:2018dyq}. In addition, we expect dark charged states at the $f_b$ scale which will appear as millicharged states in the canonical basis. These states are phenomenologically interesting for the hierarchical situation where $f_b\sim \TeV$~\cite{Izaguirre:2015eya,Berlin:2018bsc, Ball:2020dnx,Budker:2021quh}.

\section{ALP phenomenology}
\label{sec:ALP_pheno}

The results of section~\ref{sec:GUTALP} indicate that in GUTs the expectation is that ALPs will have ratio $g_{a\gamma \gamma}/m_a$ smaller than the QCD prediction -- falling to the right of the QCD line in $g_{a\gamma \gamma} - m_a$ space. The lack of experimental access to this region of parameter space highlights the need for experimental approaches which do not rely on the ALP-photon coupling. Of particular interest in this direction are the axion couplings to fermions. As can be seen from the operator
\begin{equation}\label{eq:alp_fermion}
    c_{ij}\frac{\partial_\mu a}{f_a}\bar{\psi}_i\gamma^\mu\gamma^5\psi_j\,,
\end{equation}
the fundamental difference with respect to the coupling to gauge bosons is that the fermion couplings are always shift-symmetric. Derivative couplings therefore have no topological protection and can be generated by renormalisation effects even if set to zero at a given energy scale. The phenomenology of ALPs in GUT theories is therefore similar to the photophobic~\cite{Craig:2018kne} or gluophobic~\cite{Marques-Tavares:2018cwm} ALP models, which also have suppressed couplings to gauge bosons but no suppression of fermion couplings.

%In section (\ref{sec:GUTQCDaxion}) we have shown that in the absence of mass mixing the anomaly is saturated by a single axion, meaning that all other axions have vanishing couplings to gauge bosons in first approximation. Furthermore, mass mixing only generates $m^2/m_{QCD}^2$ suppressed couplings for light ALPs. This \textit{gaugephobia} applies to all the gauge bosons coming from the unified group $\GUT$ (see \cite{Craig:2018kne} for a realisation of \textit{photophobia} or \cite{Marques-Tavares:2018cwm} for \textit{gluonphobia}) but does not have an equivalent for fermion couplings which can remain unsuppressed, $c_{ij}\sim O(1)$. 
%The reason is that no symmetry prevents the appearance of this couplings nor any topological argument forces them to be running independent. 

ALP-fermion couplings possess some model dependence as they depend on the $U(1)$ charge assignments for fermions, so in this section we briefly describe what the general expectations for these couplings are in theories with several axion fields and an underlying GUT. In field theory language, the kind of ALPs that we discuss in this section are those where the mixed $[\GUT]^2\times U(1)$ anomaly cancels but the $U(1)$ symmetry still has a chiral charge assignment. In other words, we are interested in the chiral, anomaly-free part of the group
\begin{equation}
    U(1)_{PQ}\times \Pi_iU(1)_i\,.
\end{equation}
For the sake of concreteness, let us consider a simple toy example. We assume a standard $SO(10)$ GUT with SM fermions in 3 copies of the $\mathbf{16}$ spinor. In addition to the anomalous $U(1)_{PQ}$ symmetry giving us the QCD axion, one can consider a $U(1)$ symmetry with charges +1, 0, -1 for each spinor. One trivially sees that this ALP will not couple to photons or gluons (as the $[SO(10)]^2\times U(1)$ anomaly cancels) but still couples at tree-level to fermions. This model has no special features but gives an example of how $\mathcal{O}(1)$ ALP-fermion couplings can appear.

A generic ALP coupled to fermions through (\ref{eq:alp_fermion}) has flavor conserving and flavor violating couplings at tree-level. In the case of flavor conserving couplings the strongest constraints come from astrophysical probes which, for example, can place restrictive bounds to the axion-electron coupling,  $c_{ee}/f_a \lesssim O(10^{-9})$ GeV$^{-1}$ (see~\cite{Irastorza:2018dyq} for a review). Recently the XENON1T collaboration~\cite{XENON:2020rca} has reported an excess on electron recoils that can be accommodated with a non-anomalous PQ symmetry~\cite{Takahashi:2020bpq}. If confirmed it would provide an interesting signal of physics beyond the SM, possibly connected to the flavor structure of the SM~\cite{Han:2020dwo,Han:2022iig}.

On the other hand, flavor violating couplings offer stringent constraints for the cases with off-diagonal coefficients, $c_{ij}\neq 0$~\cite{Wilczek:1982rv}. To cite some examples, the exotic decays $\mu\rightarrow e a$ and $K\rightarrow \pi a$ provide the bounds:
\begin{equation}
    \frac{c_{e\mu}}{f_a}\lesssim 1.2\times 10^{-9}\text{ GeV}^{-1}\,,\,\,\,    \frac{c_{sd}}{f_a}\lesssim 1.2\times 10^{-12}\text{ GeV}^{-1}\,.
\end{equation}
We refer the reader to \cite{MartinCamalich:2020dfe,Bauer:2021mvw} for a comprehensive analysis.
\\\\
Light bosons may also mediate long-range forces which can produce observable effects in macroscopic objects. This is the case for scalar bosons mediating monopole-monopole (in the sense of multipole expansion) interactions between fermions, and pseudoscalar fields which lead to dipole-dipole spin-dependent forces between fermions. Axions mediate a new type of interaction which has a monopole-dipole nature \cite{Moody:1984ba}. This has lead to new experimental ideas to look for axions in the laboratory~\cite{Arvanitaki:2014dfa,Raffelt:2012sp,Okawa:2021fto}. Being $\P$ and $\CP$ violating this axion-mediated monopole-dipole interaction is necessarily proportional to the topological phase $\theta_{\rm eff}$ and therefore involves axion-gauge boson couplings. 

\begin{figure*}[t]
	\centering
	\includegraphics[width=0.7\textwidth]{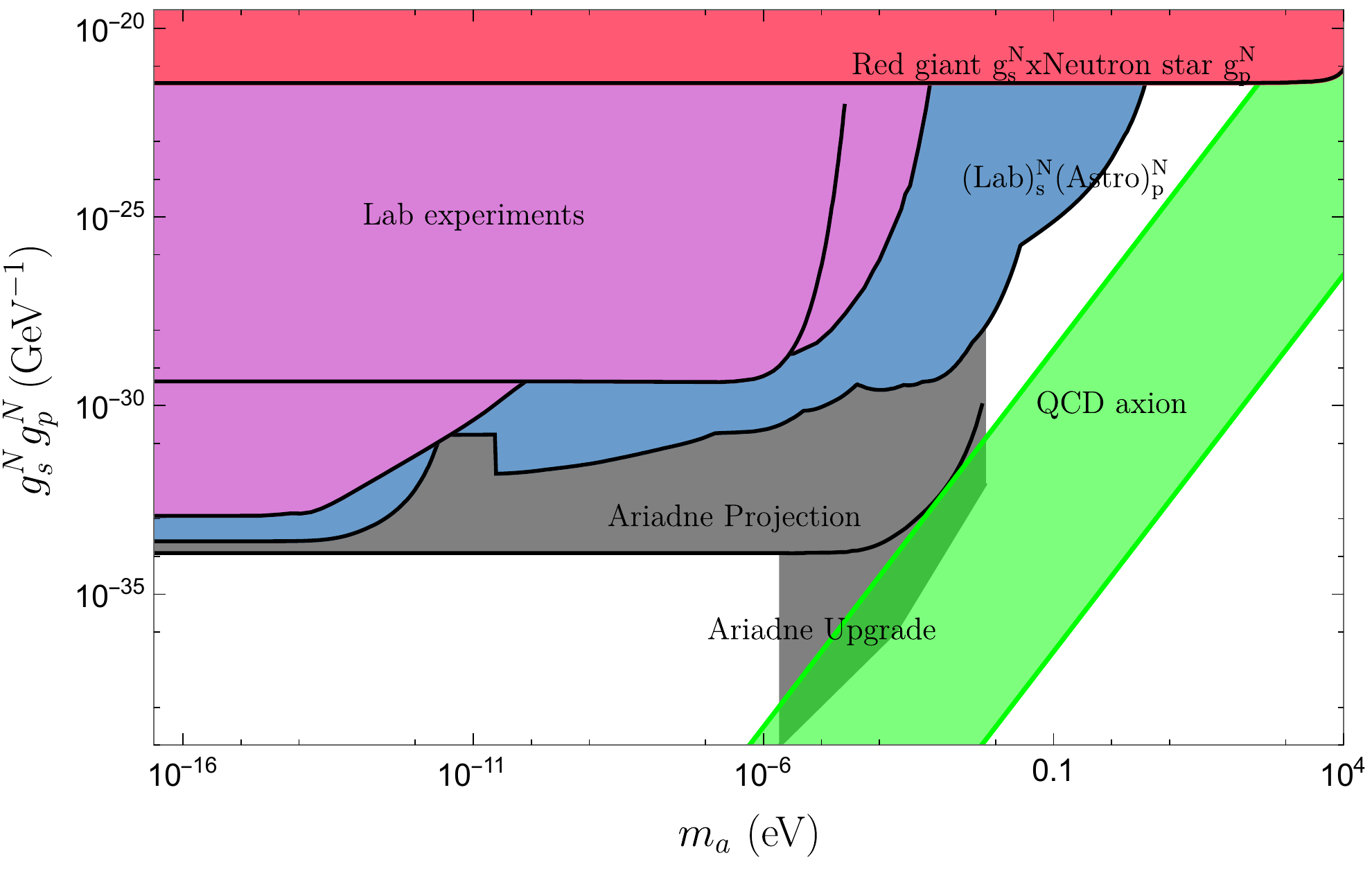}

	%\caption{\MN{\tt{placeholder, need to show experimental bounds on emergent axion}}}
		\caption{Limits on $\CP$-violating monopole-dipole interactions for nucleons, adapted from~\cite{OHare:2020wah}. Purple is different LAB exclusion limits,  blue is astro-lab exclusion, gray is the projected limits for ARIADNE. The QCD axion prediction is in green color.}

	\label{fig:axion_NN_forces}
\end{figure*}

In the GUT framework we considered in previous sections only the QCD axion and ALPs with comparable (or larger) masses will lead to sizeable monopole-dipole interactions. Additional light ALPs ($m_b \ll m_{\QCD}$) will give monopole-dipole, or monopole-monopole interactions which are generated by mixing with the QCD axion and are subsequently suppressed by $m_b^2/m_{QCD}^2$, $m_b^4/m_{QCD}^4$ respectively. For example, the scalar and dipole coupling constants that characterize the monopole-dipole interaction between nucleons for a light ALP $b$ with mass $m_b$ are roughly given by:
\begin{equation}
    g_s^N g_p^N\sim \theta_{\eff}\frac{m_q^2}{f_a^2}\frac{m_b^2}{m_{\QCD}^2}\,,
\end{equation}
with $m_q$ a light quark mass. In the same way the axion couplings to gauge bosons are highly suppressed for light axions, long-range forces other than dipole-dipole interactions are strongly screened for axions with large wavelength. Figure~\ref{fig:axion_NN_forces} shows the limits and projections for experiments looking for monopole-dipole interactions of nucleons. The QCD axion prediction is shown in green (taking $\theta_\eff$ to lie in the range $10^{-20}<\theta_\eff<10^{-10}$) and sets an upper bound for the possible magnitude of the force for a given ALP mass in GUTs. A discovery of a force mediated by an axion above the QCD band is inconsistent with unification into a simple group in the UV. Experiments looking for long range monopole-dipole forces~\cite{Arvanitaki:2014dfa, ARIADNE:2017tdd} can therefore probe GUTs in a very similar way to experiments which aim to measure the axion-photon coupling.

%\MR{We do not expect suppression for ARIADNE \cite{Arvanitaki:2014dfa} but we do for longer range forces?? \cite{Raffelt:2012sp,Okawa:2021fto}}

A final aspect of ALP phenomenology that deserves mention is related to collider physics. The region of parameter space that can be probed at these facilities -- that is, masses above the MeV scale and couplings of order $c_{ij}/f\sim O(1)$ TeV$^{-1}$ -- corresponds to the heavy axion regime and no specific prediction arises for ALP couplings in this region of parameter space. A standard analysis (see \cite{Jaeckel:2012yz,Jaeckel:2015jla}, and~\cite{Bauer:2017ris} for a recent review) applies even in the presence of a simple GUT in the UV.

\section{Non-simple groups}

\label{sec:motorbikes}

In this section we consider some well-motivated models where the SM is not embedded in a simple gauge group in the UV but some or all of the predictions of GUT theories are retained. These include the Pati-Salam model~\cite{Pati:1974yy} based on the group $SU(4)_C\times SU(2)_L \times SU(2)_R$, Trinification~\cite{Babu:1985gi} based on $SU(3)_C\times SU(3)_L\times SU(3)_R$ and flipped $SU(5)$ which is based on the gauge group $SU(5)\times U(1)_X$~\cite{Barr:1981qv,Derendinger:1983aj}. These scenarios also arise as intermediate steps in the symmetry breaking chain of more fundamental GUTs based on the gauge groups $SO(10)$~\cite{Georgi:1974sy,Fritzsch:1974nn} and $E_6$~\cite{Gursey:1975ki}, however, if the gauge group is ultimately simple in the UV then the results of sections~\ref{sec:GUTQCDaxion}~\&~\ref{sec:GUTALP} apply.

Flipped GUTs are well-motivated in certain string frameworks because they do not need adjoint fields to break the gauge group down to the SM~\cite{Antoniadis:1989zy,Font:1990uw}. Additionally, in specific set-ups the doublet-triplet splitting can also be solved~\cite{Antoniadis:1989zy}. Unlike the minimal Georgi-Glashow $SU(5)$ model, flipped $SU(5)$ also includes small neutrino mass generation as a right-handed neutrino lies inside one of the representations. This particle gets a large Majorana mass after the spontaneous gauge symmetry breaking down to the SM, giving us the possibility of implementing the seesaw mechanism. Pati-Salam and Trinification, being manifestly left-right symmetric, are well-suited to explain the SM hypercharges and relate them to the SM  baryon and lepton numbers. Assuming that the fermion content has non-exotic representations (as discussed below) this provides a way to address electric charge quantisation in integers for isolated states. Additionally, one usually links the breaking of the left-right symmetry at a high scale to the generation of small neutrino masses. 

%In this section we show that the appearance of a light ALP correlates with losing some of the GUT predictions, such as charge quantisation (integer electric charges for isolated states), the value of the weak mixing angle, and the non-existence of exotics. Even in the absence of an underlying simple gauge group, keeping some of these predictions requires that the QCD axion is the only axion with sizeable coupling to photons.

\subsection{Flipped GUTs}
A well known situation where we have a partially unified theory is the case of the flipped $SU(5)$ theory~\cite{Barr:1981qv,Derendinger:1983aj}, where the gauge group is given by $SU(5)\times U(1)_X$. The SM fermions (plus right-handed neutrinos) are contained in the representations $\mathbf{5}_{-3}$, $\bar{\mathbf{10}}_1$ and $\mathbf{1}_5$, with the subscript denoting the $U(1)_X$ charge. The scalar sector also differs from the standard $SU(5)$ model where the initial GUT gauge symmetry is broken by a Higgs in the adjoint representation, $\mathbf{24}$. Instead, in flipped $SU(5)$ the breaking down to the SM is carried by a Higgs transforming as $\mathbf{10}_1$.

After properly normalizing the generators the gauge couplings of $SU(5)\times U(1)_X$ -- that is, $\alpha_5$ and $\alpha_X$ -- obey the tree-level matching condition:
\begin{equation}
    \alpha_2(\MGUT)=\alpha_3(\MGUT)=\alpha_5(\MGUT)\,,
    \hspace{1cm}
    \frac{25}{\alpha_{1}(\MGUT)}
    =\frac{1}{\alpha_5(\MGUT)}+\frac{24}{\alpha_X(\MGUT)}\,.
\end{equation}
In flipped $SU(5)$ the weak mixing angle is given at the GUT scale by:
\begin{equation}\label{mixing_angle_flipped_SU5}
\sin^2\theta_w (\MGUT)=\frac{\alpha_Y}{\alpha_2+\alpha_Y}=\frac{3/8}{1+3/5\left ( \frac{\alpha_5}{\alpha_X}-1\right )}\,.
\end{equation}
When $\alpha_5=\alpha_X$ at the GUT scale the standard prediction $\sin^2\theta_w (\MGUT)=3/8$ is retained, which occurs naturally when the theory is embedded in a higher rank GUT. In the general case, however, the couplings are independent parameters. 

If one abandons further unification of the $U(1)_X$ factor there can exist an axion coupled to photons without receiving a QCD potential. This occurs if there are particles charged under $U(1)_X$ generating an anomaly. However, the usual GUT prediction for $\sin^2 \theta_w$ is lost. This also implies giving up the explanation of the integer quantisation of electric charges, as there may exist fractionally charged isolated states -- e.g. a fermion in the $\mathbf{1}_1$ representation would have an exotic electric charge $q=1/5$.

\subsection{Pati-Salam and Trinification}
Pati-Salam \cite{Pati:1974yy} and Trinification \cite{Babu:1985gi} are both well-known (partially) unified theories where the hypercharge comes from a diagonal generator of a non-abelian group. Both theories are manifestly left-right symmetric and are the maximal subgroups of $SO(10)$ and $E_6$, respectively. We consider the Trinification model here but similar arguments apply for Pati-Salam.

In Trinification the UV gauge group is $SU(3)_C\times SU(3)_L\times SU(3)_R$ and the electric charge generator is given by:
\begin{equation}
    Q_{\EM}=T_3^L+Y=T_3^L+T_3^R+\sqrt{\frac{1}{3}}(T^{8}_{L}+T^{8}_{R})\,.
\end{equation}
This means that, in terms of the Dynkin indices $t_r$ and the $\PQ$ charge $q_r^{\PQ}$ of each representation $r$, the ratio of anomaly coefficients can be written as:
\begin{equation}
    \frac{E}{N}%=& \frac{\text{Tr} \left[Q_{\rm PQ} \{Q_{\EM},Q_{\EM}\}\right]} {\text{Tr} \left[Q_{\rm PQ}\{T_a^{\QCD},T_b^{\QCD}\}\right]}
    = 
  %  \sum_\psi t^{(L)}_{\psi} + t^{(R)}_{\psi} q^{\PQ }_{\psi}
    (1+1/3) \left( \sum_{r \in SU(3)_{L, R}} t_r  \,  q^{\PQ}_r \right)
    / \left( \sum_{r \in SU(3)_C} t_r\,q^{\PQ}_r \right) \, .
\end{equation}
If there is a representation without color charge that is charged under $\PQ$ we may not only have $E/N\neq 8/3$, but also a light ALP coupled only to photons ($N=0$).

The prediction of the weak mixing angle can be accommodated if there exists a $\mathbb{Z}_3$ symmetry relating different gauge groups so that $\alpha_C=\alpha_L=\alpha_R$ in the UV. If all axions are singlets under the $\mathbb{Z}_3$ symmetry then the three anomaly coefficients which contribute to $E/N$ must be equal. This corresponds to the results in previous sections where one axion has $E/N = 8/3$ while the anomaly coefficients for other axions are zero. Another possibility is that each $SU(3)$ factor couples anomalously to a different axion and these three axions are permuted under the exchange symmetry in the same way as the gauge bosons. In this case there are three axions coupled to photons with equal $SU(3)$ anomaly coefficients. In some sense this resembles the GUT-charged axion scenario in section \ref{subsec:emergent}. The main difference is that since the axions are now only charged under a discrete symmetry, there is no gauge boson generating a perturbative mass.

The group structure of the trinification model with a $\mathbb{Z}_3$ is similar to the mirror matter models discussed in section~\ref{sec:discrete}, with the gauge group heuristically appearing as $G^N \rtimes \mathbb{Z}_N$. The difference between the two is that in mirror matter models the SM is embedded in a single copy of the gauge group while in trinification the different SM gauge groups come from different $SU(3)$ factors. This allows the possibility of an ALP coupled only to photons in trinification but not in the mirror world scenario.

%\MN{ $Z_3$ at GUT scale unifies coupling constants and sets decay constants and anomaly coefficients equal, gives correlation b/w axions and ALPs + add trimmed version of fractional charges discussion. Correlation b/w fractional charges and $E/N \neq 8/3$.}

Trinification can also generically have fractionally charged states.
To study the absence of fractional charges it is useful to work with the charge~\cite{Polchinski:1998rr}:
\begin{equation}\label{Q'_charge}
    Q'=Q_{\EM}+\frac{T_{\rm color}}{3}\,.
\end{equation}
The operator $T_{\rm color}$ denotes the $SU(3)_C$ colour triality of a representation; confinement implies that all isolated states are color singlets with $T_{\rm color}=0$ mod 3. Even though quarks have a fractional $Q_\EM$ they have an integer $Q'$ charge, so that for hadrons both $Q'$ and $Q_{\EM}$ are integers. 
%The value of triality for each representation is fixed (for example triplets contribute with $+1$, and sextets with $+2$) with conjugate representations contributing with a (-) sign. 
The requirement of having only integer charges determines the allowed matter representations. Together with the $\mathbb{Z}_3$ symmetry this turns out to be quite constraining and the only possible set of representations that is chiral, anomaly free with integer electric charges is:
\begin{equation}
    (\mathbf{3},\bar{\mathbf{3}},1)+(\bar{\mathbf{3}},1,\mathbf{3})+(1,\mathbf{3},\bar{\mathbf{3}})\,, 
\end{equation}
plus other representations obtained by taking tensor products of this set. For example, additional representations of the form $(\mathbf{3},1,1)+(1,\mathbf{3},1)+(1,1,\mathbf{3})$ will lead to fractionally charged hadrons or leptons.  If particles in these representations exist at all they must be well above any explored energy scale in order to satisfy collider and cosmological bounds~\cite{Perl:2009zz}.

\section{Conclusion}
\label{sec:conclusion}

Topological interactions are unique in that they are largely unchanged from the UV to the IR, offering a way to test the far UV dynamics of theories using low-energy experiments. The axion-photon coupling is an example of a topological quantity that is highly relevant for the large experimental program being conducted to search for axions. In this work we have studied in detail the restrictions on this coupling from the requirement that the SM is unified into a simple gauge group at a fundamental scale. This requirement means that any axion with a coupling to photons must also couple with a comparable strength to gluons and therefore receives a contribution to its mass from QCD instantons.

In the absence of mixing effects this implies that in a simple GUT theory there is only one axion with an anomalous coupling to both gluons and photons and corresponds to the well-studied case of the QCD axion. Once mixing effects are considered there may be other ALPs which couple to photons, but the restrictions imposed by unification in the UV imply that they have a coupling to mass ratio smaller than that of the QCD axion. If axions are charged under the GUT gauge group, they can couple preferentially to photons, but pick up a mass from GUT interactions. If the axions emerge as composite states their mass can be much lower than the GUT scale due to compositeness. In this case the ALP coupling to photons has a different dependence on the ALP mass and current experimental bounds require that $m_a \gtrsim 1 \GeV$ and $f_a > 10^5$ TeV. %\textcolor{red}{These composite ALPs might be accessible in collider experiments.}

There are also models which preserve some of the phenomenological predictions of GUT theories without unifying the SM into a simple gauge group in the UV. These theories were studied in section~\ref{sec:motorbikes}. Preserving the theoretical predictions of GUTs -- coupling unification, predicting the weak-mixing angle and the absence of fractionally charged states -- is still somewhat correlated with ALP phenomenology, but the details depend on the model. 

These results mean that the discovery of a light ALP with an
observable photon coupling is not consistent with a simple GUT UV
completion of the SM. Low-energy searches for light ALPs through the
ALP-photon coupling, cosmological signatures such as the rotation of
CMB polarisation by light axions coupled to photons
\cite{Lue:1998mq,Arvanitaki:2009fg} and their associated strings
\cite{Agrawal:2019lkr}, and long-range monopole-dipole forces are
therefore a novel way to test if the SM is unified at a fundamental
level. A recent analysis of \textit{Planck} and \textit{WMAP} data has
reported a hint of a non-zero cosmic birefringence angle
$\beta={0.342^\circ}^{+0.094^\circ}_{-0.091^\circ}$, excluding $\beta=0$ at $3.6\sigma$ \cite{Minami:2020odp,Eskilt:2022cff}. This signal seems to be consistent with an ALP coupled to photons with a mass $H_{\text{CMB}}>m_a >H_0$. The fate of this signal will be decided in future CMB experiments. 
ALP-fermion derivative couplings are not topological quantities so are not suppressed in the same way as the ALP-photon couplings. Searches for ALPs through these interactions offer a promising approach to search for light ALPs in GUT theories and can provide the strongest constraints on these models in the regime where the ALP-photon coupling is small.

It is intriguing that GUT theories defined in the far UV provide strong restrictions in IR physics with experimental implications. As we have shown in this work, ALP searches are a low-energy probe of grand unification just like the classic signature of proton decay. %ALP searches deserve to be included in the catalogue of experimental probes of Grand Unification.

\iffalse

\section{Useful papers}

\MN{Some more ref's - have been going through places where we cite in the intro to try and see if there are any we may have missed. Not sure if all these are relevant but anyway:
\begin{itemize}
    \item Experimental reviews - \cite{Essig:2013lka}
    \item String theory axions - \cite{Witten:1984dg,Conlon:2006tq}
    \item PQ/QCD axion - \cite{Peccei:1977ur}
    \item ALPs/ALP searches - \cite{Masso:1995tw, Ringwald:2012hr}
    \item Quality problem/axion mass in quantum gravity - \cite{Holman:1992us,Kamionkowski:1992mf,Kallosh:1995hi}
    \item Flavour test of axion/ALP couplings - \cite{Izaguirre:2016dfi}
    %\item GUT axion models w/ $f_a = M_{GUT}$ - \cite{Georgi:1981pu,Nilles:1981py} 
  %  \item axiverse - \cite{Cicoli:2012sz}
    \item ALP searches (refs all stolen from \cite{Bauer:2017ris} \& \cite{Izaguirre:2016dfi}) - \cite{CHARM:1985anb, Riordan:1987aw,Bjorken:1988as, Blumlein:1990ay,OPAL:2002vhf, Jaeckel:2012yz, ATLAS:2014jdv, Mimasu:2014nea, ATLAS:2015rsn, Jaeckel:2015jla, Dobrich:2015jyk, Alekhin:2015byh,Kleban:2005rj}
\end{itemize}}

\fi

\section*{Acknowledgments}

We thank  Martin Bauer, Cliff Cheung, Djuna Croon, Fran Day, Rachel Houtz, David Kaplan, Gustavo Marques-Tavares,  Clara Murgui, Surjeet Rajendran, Raman Sundrum and Mark Wise for useful discussions. We also thank JiJi Fan, Anson Hook, Junwu Huang, John March-Russell, and Matt Reece for enlightening conversations and comments on the manuscript. We also thank \textcolor{orange}{Arthur Platschorre} for taking us to the Tree Artisan Cafe.
PA is supported by the STFC under Grant No. ST/T000864/1. MN is funded by a joint Clarendon and Sloane-Robinson scholarship from Oxford University and Keble college.

\appendix

\section{Level of embedding}

\label{sec:embedding}

%\textcolor{red}{JMR suggests to put this earlier in the paper... I prefer to keep it as a separated appendix so that people that want to skip technical details associated to the embedding can go to the simple case and read the paper. Those interested in the more general thing can come here and read the details...}

It is commonly assumed in the literature that the level of embedding of the different SM groups are level 1. This means that the SM descends from an $SU(5)$ subgroup of the possibly larger, simple gauge group unifying interactions in the UV. While this is a compelling choice it is not fully general. Indeed, when the gauge group is sufficiently large there can be different non-trivial ways to obtain the SM at low energies, however it is very hard to find any concrete examples with a consistent low-energy spectrum.

A simple example illustrating a higher-level embedding is where one of the non-Abelian groups of the SM arises as a diagonal subgroup of a $k$-product group:
\begin{equation}\label{k-level-SSB}
    G_1\times ... \times G_k\rightarrow G_{\rm diag} \,,
\end{equation}
corresponding to the level of embedding of $G_{\rm diag}$ being $k$. The generators of $G_{\rm diag}$ are given as a linear combination of the original generators, $T^a=\sum^{k}_{i=1} T^a_i$, and the tree-level matching condition on the gauge coupling $\frac{1}{\alpha}=\sum^{k}_{i=1} \frac{1}{\alpha_i}$ has to be satisfied at the SSB scale. One can also obtain higher-level embeddings by using gauge groups with non simply-laced algebras -- that is, those which contain roots of different lengths~\cite{Dienes:1996yh}.

In general, any possible embedding of the SM in a GUT can be labeled by 2 integers $(k_2,k_3)$ describing the level of embedding of the $SU(2)$ and $SU(3)$ sectors and the hypercharge normalisation $k_1$, which is a rational number. The regular embeddings in $SU(5)$~\cite{Georgi:1974sy}, $SO(10)$~\cite{Georgi:1974sy,Fritzsch:1974nn} and $E_6$~\cite{Gursey:1975ki} correspond to $(k_2,k_3;k_1)=(1,1;5/3)$. Several of the GUT predictions depend on these quantized numbers. For example, the weak mixing angle prediction at the unification scale is in general not fixed and depends on $k_1$, $k_2$ as~\cite{Font:1990uw}
\begin{equation}
    \sin^2\theta_w=\frac{k_2}{k_1+k_2}\,,
\end{equation}
and the ratio of anomaly coefficients for axion-photon coupling is given by
\begin{equation}
    \frac{E}{N}=\frac{k_2+k_1}{k_3}\,.
\end{equation}
The value of $E/N$ coincides with the inverse of $\sin^2\theta_w$ at the unification scale if $k_2=k_3$.  We see that in the well-studied GUT models with standard embedding -- e.g.~$SU(5)$ with $\mathbf{\bar 5} + \mathbf{10}$, $SO(10)$ with $\mathbf{16}$ and $E_6$ with $\mathbf{27}$ -- 
we get $(k_2,k_3;k_1)=(1,1,5/3)$, and $E/N=8/3$. While this appears to be not the most general possibility, we highlight below the severe model building challenges to find a different embedding of the SM in a 4D simple gauge group.

In higher embeddings, the action of UV instantons is reduced compared to the standard embedding~\cite{Agrawal:2017ksf,Fuentes-Martin:2019bue,Csaki:2019vte}. 
\begin{align}
    S_k\sim \frac{S_{k=1}}{k} 
    \sim 
    \frac{2\pi}{ \aGUT k}\,.
\end{align}
Small size gauge instantons (SSI) at the scale of symmetry breaking are not aligned with the QCD axion potential in general, and may reintroduce the strong $\CP$ problem for a higher level embedding for  $k$ as low as 2~\cite{Svrcek:2006yi}.

Perhaps more importantly, chiral exotics and fractionally charged states are hard to avoid for generic choices of $(k_2,k_3;k_1)$. 
%\MR{I have a kind of proof that in 4dimensions $k_2=k_3$ is the only possibility unless one has chiral exotics or gauge anomalies. Not only that, it seems that the only possibilities are  $k_2=k_3=1$ or 3}
As an example, let us consider the  k-level embedding of $SU(N)$ in a diagonal subgroup as above (similar arguments apply for other groups):
\begin{equation}
    SU(kN)\rightarrow [SU(N)]^k\rightarrow SU(N)_\text{diag}\,.
\end{equation}
The fundamental of $SU(kN)$ has the following branching rule:
\begin{equation}
    \textbf{kN}\rightarrow (\textbf{N},\textbf{1},...)+ ... + (\textbf{1},..,\textbf{N})\rightarrow k \times \textbf{N}\,.
\end{equation}
yielding $k$ copies of the fundamental of $SU(N)_\text{diag}$. Any other representation -- and its branching rules -- can be obtained by taking tensor products of the fundamental. This has the important implication that given an anomaly-free, chiral (complex) set of $SU(kN)$ representations, any complex representation $R$ of $SU(N)_\textsf{diag}$ will come in a number given by  $mk$, where $m$ is  an integer (in general we expect different $m$ for different representations, $R$). Since the set of representations was complex with respect to $SU(kN)$, this means that in general we  expect to have 
\begin{equation}
mk \text{ copies of } R\, \,+\,\, \tilde{m} k \text{ copies of } \bar{R}\,,
\end{equation}
where $\bar{R}$ is the complex conjugate of $R$. 

In terms of fermions, each $\bar{R}$ can pair up with a $R$ and get a large vector-like mass. Thus, at low energies, we expect to have a number of chiral fermions in the $R$ representation of $SU(N)_\text{diag}$ given by:
\begin{equation}
    (m-\tilde{m})k\,,
\end{equation}
For example, if $SU(2)_L$ is embedded at level $k_2$, we expect at least $k_2$ lepton doublets \emph{and} $k_2$ quark doublets (each representation with a different $SU(3)$ or $U(1)$ charge is a different $R$). Since a chiral leptonic 4th family is excluded, we find the bound $k_2\leq 3$. In addition, if $k_2=2$, we expect lepton doublets to come in multiples of 2, which is either too few or too many families. The only possibilities are:
\begin{equation}
    k_2=1\,, \text{ and }\, k_2=3\,.
\end{equation}

We can now use anomaly cancellation -- which in the SM occurs family by family -- to argue that we need the same number of lepton doublets as quark doublets. Not only that, to avoid $[SU(3)_C]^3$ anomalies we need the same number of triplets as anti-triplets. Therefore, in the absence of chiral exotics, it is easy to conclude that we get only 2 consistent possibilities:
\begin{equation}
    k_2=k_3=1\,, \text{ and }\, k_2=k_3=3\,.
\end{equation}
The criterion of anomaly cancellation does not give any hint about what the hypercharge normalisation is, as it is just a global factor in the anomalies that involve $U(1)$. However, from the definition of the electric charge operator, we find that in a situation where $k_1$ does not satisfy $k_1/k_2 = 5/3$, we expect isolated states with electric charge smaller than the electron charge in the spectrum.

While this is by no means a proof that generic embeddings are not theoretically consistent, it highlights the serious model building challenges that come with such non-standard situations. This suggests the striking result that $k_2=k_3=1$ or $k_2=k_3=3$, which taking a right value for hypercharge normalisation gives
\begin{equation}
    \frac{E}{N}=\frac{8}{3}\,.
\end{equation}
It will be nice to build realistic 4 dimensional GUT models where the index of embedding is related to the replication of families.

\bibliographystyle{utphys}
\bibliography{newrefs_axion}

\end{document}